\documentclass[aps,prx,amsmath,amssymb,reprint]{revtex4-2}

\usepackage[utf8]{inputenc}
\usepackage[T1]{fontenc}
\usepackage{graphicx}
\usepackage{bm}
\usepackage{xcolor}
\usepackage[normalem]{ulem}
\usepackage[hidelinks]{hyperref}

\renewcommand*{\vec}[1]{\mathbf{#1}}
\newcommand*{\diff}[1]{\mathrm{d}#1\,}
\newcommand{\rv}{{\mathbf r}}
\newcommand{\pv}{{\mathbf p}}
\newcommand{\rmexc}{{\rm exc}}
\newcommand{\rmext}{{\rm ext}}
\newcommand{\rmid}{{\rm id}}

\graphicspath{{./figures}}

\begin{document}

\title{Neural density functional theory of liquid-gas phase coexistence}

\author{Florian Sammüller}
\email{florian.sammueller@uni-bayreuth.de}
\author{Matthias Schmidt}
\email{matthias.schmidt@uni-bayreuth.de}
\affiliation{Theoretische Physik II, Physikalisches Institut, Universität Bayreuth, 95447 Bayreuth, Germany}
\author{Robert Evans}
\affiliation{H.~H.~Wills Physics Laboratory, University of Bristol, Royal Fort, Bristol BS8 1TL, United Kingdom}

\date{\today}

\begin{abstract}
We use supervised machine learning together with the concepts of classical density functional theory to investigate the effects of interparticle attraction on the pair structure, thermodynamics, bulk liquid-gas coexistence, and associated interfacial phenomena in many-body systems.
Local learning of the one-body direct correlation functional is based on Monte Carlo simulations of inhomogeneous systems with randomized thermodynamic conditions, randomized planar shapes of the external potential, and randomized box sizes.
Focusing on the prototypical Lennard-Jones system, we test predictions of the resulting neural attractive density functional across a broad spectrum of physical behavior associated with liquid-gas phase coexistence in bulk and at interfaces.
We analyse the bulk radial distribution function $g(r)$ obtained from automatic differentiation and the Ornstein-Zernike route and determine i) the Fisher-Widom line, i.e.\ the crossover of the asymptotic (large distance) decay of $g(r)$ from monotonic to oscillatory, ii) the (Widom) line of maximal correlation length, iii) the line of maximal isothermal compressibility and iv) the spinodal by calculating the poles of the structure factor in the complex plane.
The bulk binodal and the density profile of the free liquid-gas interface are obtained from density functional minimization and the corresponding surface tension from functional line integration.
We also show that the neural functional describes accurately the phenomena of drying at a hard wall and of capillary evaporation for a liquid confined in a slit pore.
Our neural framework yields results that improve significantly upon standard mean-field treatments of interparticle attraction.
Comparison with independent simulation results demonstrates a consistent picture of phase separation even when restricting the training to supercritical states only.
We argue that phase coexistence and its associated signatures can be discovered as emerging phenomena via functional mappings and educated extrapolation.
\end{abstract}

\maketitle

\section{Introduction}

The emergence of two or more distinct thermodynamic stable phases on varying thermodynamic conditions is arguably one of the most striking phenomena in statistical mechanics, whether this occurs in bulk or at interfaces, in pure or in multi-component systems.
Unsurprisingly, the recent surge in the use of machine-learning techniques in physics \cite{Carrasquilla2017,Bedolla2020} has focused on the prototypical (lattice-based) Ising model for developing appropriate techniques and strategies to investigate phase-separating systems.
Such work includes the finite-size analysis in neural network classification of critical phenomena \cite{Chertenkov2023} and mapping out phase diagrams with generative classifiers \cite{Arnold2024}.
Detecting the approach to a critical point also constitutes a central task in more general dynamical systems, which commonly requires the application of advanced computational methods \cite{Liu2024,Harris2024,Morr2024}.

In this paper we focus on using machine learning to investigate liquid-gas phase separation and related phenomena in a \emph{continuum} model fluid, namely the Lennard-Jones system, arguing that understanding the physics in such a simple model, which encompasses both repulsive and attractive interparticle interactions, provides a basis for understanding the occurrence of the same phenomena that arise in more complex fluids \cite{Hansen2013, Frenkel2023}.

We choose to employ the formal techniques of classical density functional theory (DFT) \cite{Evans1979, Evans1992, Evans2016, Hansen2013} to investigate both bulk and inhomogeneous (interfacial) properties.
Going from a bulk (homogeneous) system with constant density to the locally resolved one-body density profile $\rho(\rv)$ of an inhomogeneous fluid, where $\rv$ indicates position, allows one to formulate the statistical mechanics based on functional relationships, as described briefly below.
Although the DFT framework is formally exact, approximations are required in implementations.
The conventional approach is to treat short-ranged repulsion in terms of a hard-sphere free energy functional, e.g.\ via fundamental-measure theory \cite{Rosenfeld1989, Roth2010}, and the attractive interaction via a simple mean-field factorization ansatz, see Ref.~\cite{Evans1992} and recent papers \cite{Evans2019, Coe2022, Coe2023}.
Such a treatment, which is in the spirit of van der Waals, generates an explicit (analytical) formula for the excess (over ideal gas) intrinsic Helmholtz free energy functional that incorporates information about the effects of repulsion and attraction.

As DFT is an exact formulation of the many-body statistical mechanics, the choice of the excess free energy functional is the only approximation that enters a given study.
Once it is specified consistent and complete investigation of a wide variety of properties can be made.
Computational limitations, leaving aside some intricacies of implementing nonlocal treatments for hard spheres \cite{Rosenfeld1989, Roth2010}, are minor, certainly for systems which exhibit planar or spherical symmetry.
Hence wide parameter sweeps and close monitoring of the effects of small changes in control parameters, especially near phase coexistence, are readily carried out enabling the investigation of subtle phenomena such as phase transitions at substrates and in confinement.

However, despite these virtues, making \emph{direct} quantitative comparisons with data from many-body simulations is often not straightforward.
Using the concept of corresponding states, i.e.\ scaling the bare thermodynamic parameters by their corresponding values at the bulk critical point, can enable meaningful comparisons but this is a pragmatic approach.
In essence, one often performs separate simulations and theoretical calculations and from their combination attempts to gather, a posteriori, a complete picture of the physics.
Finding an accurate, versatile and computationally manageable improvement on the standard mean-field treatment of attraction continues to pose a significant challenge in classical DFT.

Machine learning provides a very different perspective for addressing the limitations of analytical approaches (meaning writing down explicit free energy functionals) by incorporating quasiexact simulation reference data for the construction of functional relationships, as was pursued for the classical \cite{Lin2019, Lin2020, Cats2021, Yatsyshin2022, MalpicaMorales2023, Dijkman2024, Heras2023, Sammueller2023neural, Sammueller2024why, Sammueller2024hyper, Sammueller2024pair, Zimmermann2024, Simon2024patchy, Simon2024review, Stierle2024, Kampa2024, Yang2024} and quantum (electronic) worlds \cite{Nagai2018, Schmidt2019, Zhou2019, Nagai2020, Li2021, Li2022, Pederson2022, Gedeon2021}.
The review by \citet{Simon2024review} gives a valuable overview of very recent applications of machine-learning techniques within classical DFT and addresses methodological connections to electronic DFT and to nonequilibrium systems.
Amongst the different approaches \cite{Lin2019, Lin2020, Cats2021, Yatsyshin2022, MalpicaMorales2023, Dijkman2024, Heras2023, Sammueller2023neural, Sammueller2024why, Sammueller2024hyper, Sammueller2024pair, Zimmermann2024, Simon2024patchy, Simon2024review, Stierle2024, Nagai2018, Schmidt2019, Zhou2019, Nagai2020, Li2021, Li2022, Pederson2022, Gedeon2021, Kelley2024, Kampa2024, Yang2024} that were put forward, the neural functional theory based on \emph{local one-body learning} \cite{Heras2023, Sammueller2023neural, Sammueller2024why, Sammueller2024hyper, Sammueller2024pair, Zimmermann2024, Kampa2024} in inhomogeneous training systems has proved to be a versatile and highly accurate tool, both in equilibrium \cite{Sammueller2023neural, Sammueller2024why, Sammueller2024hyper, Sammueller2024pair, Kampa2024} as well as for microscopically resolved nonequilibrium flow problems \cite{Heras2023, Zimmermann2024}.
In the latter case the required functional relationships are those of power functional theory \cite{Schmidt2013, Schmidt2022}.

The initial appeal of the local learning approach stems from its simplicity.
In equilibrium, sampling the one-body density profile in spatially inhomogeneous systems is all that is required for the generation of a training data set.
A neural network with a simple multilayer perceptron architecture is then trained to represent the functional relationship from the density profile to the one-body direct correlation function.
The latter object is directly accessible from the input simulation data.
Moreover, it also arises as a central functional mapping in DFT; further details and comments on its significance are given below.
In particular, the short-ranged nature of direct correlation functions permits one to consider the functional mapping \emph{locally}, thereby aiding the training procedure and making efficient use of the input data.
Although applications of the resulting neural functional are rather straightforward in practical terms, these are powerful and provide access to a multitude of physical properties by making use of the underlying formal structure of DFT and liquid state theory.
These include the prediction of density profiles, also in multiscale settings \cite{Sammueller2023neural}, employing automatic differentiation \cite{Baydin2018,Sammueller2024why,Stierle2024} and numerical functional integration to determine correlation functions and thermodynamical properties, and using exact statistical mechanical sum rules \cite{Baus1984,Evans1990,Henderson1992}, specifically those that follow from Noether's theorem \cite{Hermann2021,Hermann2022,Sammueller2023noether,Hermann2024,Robitschko2024,Mueller2024}, to examine self-consistency.

The neural functional approach was investigated in great detail for models with purely repulsive (hard core) potentials.
It provides an excellent approximation for Percus' exact free energy functional for hard rods in one dimension \cite{Sammueller2024why} and was shown to constitute a clear improvement \cite{Sammueller2023neural} over the already highly accurate White-Bear Mk.\ II version of fundamental-measure theory \cite{HansenGoos2006,Roth2010} for hard spheres in three dimensions.
Importantly, the neural functional method is not restricted to hard cores; it applies to general interatomic potentials which may also include attraction.
That the neural functional can treat attraction successfully was illustrated in determinations of the structure of the Lennard-Jones fluid at a fixed supercritical temperature \cite{Sammueller2023neural, Sammueller2024pair}; this constitutes an important test case \cite{Dijkman2024}.

These recent investigations did not address the fundamental issue of how the presence of a phase transition might be accounted for within the framework of a neural density functional.
Here we focus on the liquid-gas transition which is a basic manifestation of the presence of interparticle attraction and seek to assess whether the neural functional can describe i) phase coexistence and the approach to the associated critical point, ii) surface tension and density profiles of the liquid-gas interface, iii) drying and capillary evaporation transitions that occur at subcritical temperatures and iv) how \emph{accurately} the approach performs for both bulk and interfacial properties.
To the best of our knowledge, none of these issues were addressed in previous machine-learning investigations.
Specifically, we extend the neural methodology \cite{Sammueller2023neural, Sammueller2024why} by introducing thermal training and investigate whether neural functionals can describe physical phenomena that occur in the three-dimensional Lennard-Jones system in sub- as well as supercritical regions of the phase diagram.
The answer is emphatically: yes.

The paper is organized as follows.
We describe our methodology of working with machine learning within a rigorous statistical mechanical framework in Sec.~\ref{sec:method}.
This includes a summary of the formally exact density functional foundation in Sec.~\ref{sec:backgroundDFT} and a description of the simulation-based generation of training data in Sec.~\ref{sec:data_generation}.
Details of the neural network and of the supervised training procedures are given in Sec.~\ref{sec:training}.
The resulting neural density functional theory together with the associated methods of functional calculus are laid out in Sec.~\ref{sec:neural_theory}.

All subsequent results originate from this neural functional method and are described in Sec.~\ref{sec:results}.
An account of the emerging bulk pair correlation structure is given in Sec.~\ref{sec:twobody_bulk}.
Results for the lines in the phase diagram where the isothermal compressibility and the true correlation length are maximal, together with the Fisher-Widom line and the spinodal, are presented Sec.~\ref{sec:map_T_rho}.
Our results for the liquid-gas binodal and estimate of the critical point are described in Sec.~\ref{sec:coexistence}.
The bulk equation of state, liquid-gas density profiles, and the corresponding surface tension are laid out in Sec.~\ref{sec:eos_interface_tension}.
A description of the divergence of the correlation length in the critical region and corresponding Ornstein-Zernike plots are given in Sec.~\ref{sec:critical_region}.
Results for inhomogeneous fluids, that describe our predictions for drying at a hard wall, for capillary evaporation in a slit pore and for the corresponding behavior of locally resolved density fluctuations, are presented in Sec.~\ref{sec:drying}.
In Sec.~\ref{sec:supercritical} we compare with results obtained from a neural functional trained with data from supercritical states only.
Remarkably this procedure also predicts liquid-gas coexistence and associated phenomena.

We conclude with a discussion in Sec.~\ref{sec:discussion}.
This includes an assessment of the strengths of the neural functional methodology in Sec.~\ref{sec:discussion_methodology}, an overview of the physical phenomena that we investigated and what remains to be ascertained in Sec.~\ref{sec:discussion_physics}, some speculations on the extent to which the prediction and discovery of phase coexistence can be based on functional mappings and their extrapolation in Sec.~\ref{sec:discussion_extrapolation}, and an outlook on possible future work in Sec.~\ref{sec:discussion_outlook}.

\section{Method}
\label{sec:method}

\subsection{Overview of classical density functional theory}
\label{sec:backgroundDFT}

We briefly sketch the essentials of classical DFT as a method to treat the statistical mechanics of many-body systems.
The system itself is defined by its Hamiltonian $H=\sum_i \pv_i^2/(2m) + u(\rv_1,\ldots, \rv_N) + \sum_i V_\rmext(\rv_i)$, where the sums over $i$ run over all $N$ particles, $\pv_i$ and $\rv_i$ are the momentum and position of particle $i=1,\ldots,N$ in $d$ dimensions, $m$ denotes the particle mass, $u(\rv_1,\ldots,\rv_N)$ is the interparticle interaction potential,
and $V_\rmext(\rv)$ is an external one-body potential.
The thermodynamic control parameters are the temperature $T$ and the chemical potential $\mu$ when working in the grand ensemble.
The associated thermodynamic potential is the grand potential (or grand canonical free energy), which is given as $\Omega_0(T,\mu)=-k_BT\ln \Xi(T,\mu)$, where $\Xi(T,\mu)$ is the grand partition sum and $k_B$ is the Boltzmann constant.
For compactness of notation we have suppressed the dependence on the system volume $V$.

Classical DFT \cite{Hansen2013,Evans1979,Evans1992} ascertains the existence and uniqueness of the grand potential density functional, $\Omega([\rho], T,\mu)$, which consists of ideal, excess (over ideal gas), external, and chemical contributions according to the sum
\begin{equation}
    \label{eq:omegaFunctional}
    \begin{split}
        \Omega([\rho], T,\mu) &= F_{\rm id}([\rho],T) + F_{\rm exc}([\rho],T)\\
        &\quad +\int\!\diff{\vec{r}} \rho(\rv) [V_\rmext(\rv)-\mu],
    \end{split}
\end{equation}
where the position integrals run over the system volume $V$.
Here and throughout we indicate functional relationships by square brackets.
The ideal gas free energy functional is known exactly as $F_\rmid([\rho],T)=k_BT \int \diff{\rv} \rho(\rv)[\ln(\rho(\rv)\Lambda^d)-1]$, where $\Lambda$ is the thermal de Broglie wavelength (which we will set to the particle size below).
The excess free energy functional $F_\rmexc([\rho],T)$ accounts for the effects of the nonvanishing interparticle interactions.
Crucially, instead of operating only on the true equilibrium density profile $\rho_0(\rv)$, the grand potential density functional \eqref{eq:omegaFunctional} accepts any general ``test'' function profile $\rho(\rv)$ that does not need to have particular physical significance for the system at hand.
Identifying the true equilibrium density profile $\rho_0(\rv)$ is ensured by the formally exact minimization principle $\Omega([\rho_0], T,\mu) \leq \Omega([\rho], T,\mu)$ and hence
\begin{equation}
    \label{eq:omegaMinimization}
    \frac{\delta \Omega([\rho], T, \mu)}{\delta\rho(\rv)} \Big|_{\rho=\rho_0} = 0 \qquad {\rm (min)}.
\end{equation}
Here $\delta/\delta\rho(\rv)$ indicates the functional derivative with respect to the test function $\rho(\rv)$ and the result of the differentiation is evaluated at the equilibrium density profile, $\rho(\rv)=\rho_0(\rv)$, as indicated in the notation (brief accounts of functional differentation are available \cite{Schmidt2022, Sammueller2024why}).
Furthermore, the value of the grand potential is obtained by evaluating the grand potential density functional at the equilibrium density profile:
\begin{equation}
    \Omega_0(T,\mu) = \Omega([\rho_0], T,\mu).
\end{equation}
Access to $\Omega_0(T,\mu)$ provides, in principle, full thermodynamic information, including knowledge of the phase diagram.
(In the subsequent sections, for notational simplicity, we drop the label 0 as an indicator for equilibrium.)
Despite operating entirely on the one-body level of correlation functions, in principle \emph{all} higher-body correlation functions are accessible.
While this information can come from the test-particle limit \cite{Percus1962} and hyperdensity functional concepts \cite{Sammueller2024hyper}, the standard route is via functional differentiation and the Ornstein-Zernike relation, as we sketch and use below.

The above framework is formally exact.
Having an exact form of the excess free energy functional $F_\rmexc([\rho], T)$ in Eq.~\eqref{eq:omegaFunctional}, as is available for very few one-dimensional systems \cite{Percus1976, Sammueller2024why}, then merely requires solution of the minimization problem \eqref{eq:omegaMinimization}, which is typically performed numerically (as described below in the current neural context).
No approximation has entered at this point and in principle the exact statistical mechanics is retained.

In practice, approximations are required in order to treat the nontrivial effects of interparticle interactions.
For the common case of liquid-gas phase separating systems where particles interact via a pair potential $\phi(r)$ with interparticle distance $r$, such that the total interparticle interaction energy is given by $u(\rv_1,\ldots, \rv_N)=\sum_{ij(i\neq j)}\phi(|\rv_i-\rv_j|)/2$, the standard mean-field approximation consists of assuming the splitting
\begin{equation}
    \label{eq:FexcMeanField}
    \begin{split}
        F_\rmexc([\rho], T) &= F_{\rm hs}([\rho], T)\\
        &\quad + \frac{1}{2} \int\!\diff{\vec{r}}\!\int\!\diff{\vec{r}'} \rho(\rv) \rho(\rv') \phi_{\rm attr}(|\rv-\rv'|).
    \end{split}
\end{equation}
The repulsive part of the interparticle potential, which gives rise to packing effects, is treated in terms of the hard sphere reference functional $F_{\rm hs}([\rho], T)$ in Eq.~\eqref{eq:FexcMeanField}.
This depends linearly on temperature and is typically represented by fundamental-measure theory (FMT) \cite{Rosenfeld1989, Roth2010}.
The longer-ranged, attractive part of the potential, $\phi_{\rm attr}(r)$, needs to be split off from (and continued into the core of) the full pair potential $\phi(r)$.
The approximation~\eqref{eq:FexcMeanField} provides the essential ingredients for successful competition of entropy (first term) and energy (second term) to drive bulk liquid-gas and certain surface phase transitions.
The dependence on temperature $T$ remains simplistic: linear variation (first term) on top of a constant (second term).
In bulk, Eq.~\eqref{eq:FexcMeanField} yields a generalization of the van der Waals equation of state.

Instead of working with the mean-field approximation~\eqref{eq:FexcMeanField} here we rather use machine-learning methods \cite{Sammueller2023neural, Sammueller2024why} to represent simultaneously both the packing and the attraction effects of the interparticle interactions.
While the excess free energy functional $F_\rmexc([\rho], T)$ itself can be trained on the basis of inhomogeneous local learning \cite{Sammueller2024pair}, we choose to start with the one-body direct correlation functional
\begin{equation}
    \label{eq:c1_from_Fexc}
    c_1(\vec{r}; [\rho], T) = - \frac{\delta \beta F_\mathrm{exc}([\rho], T)}{\delta \rho(\vec{r})},
\end{equation}
where $\beta = 1 / (k_B T)$ is the inverse temperature.
As we show in Sec.~\ref{sec:neural_theory}, $c_1(\vec{r}; [\rho], T)$ is directly relevant for solving the minimization problem~\eqref{eq:omegaMinimization}, but it also provides access to further physical quantities by utilization of functional calculus, in particular implemented via automatic differentiation \cite{Baydin2018, Sammueller2024why, Stierle2024}.
The nontrivial information for training $c_1(\rv; [\rho],T)$ is straightforward to access in many-body simulations, as we lay out in the following.

\begin{figure*}[tbp]
    \centering
    \includegraphics[width=\linewidth]{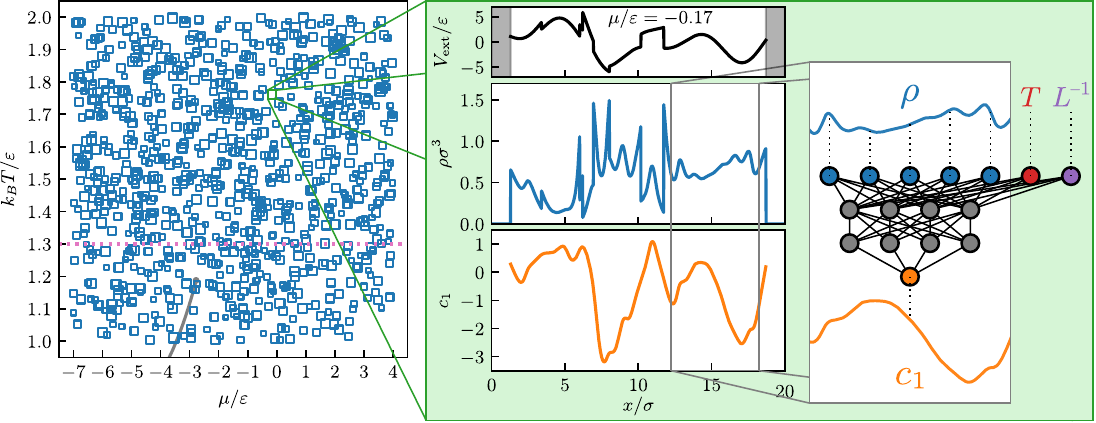}
    \caption{
        Left panel: training data are acquired via grand canonical MC simulations of the truncated LJ fluid in inhomogeneous planar environments at randomized chemical potential $\mu$ and temperature $T$.
        We show the thermodynamic state points of all contributing simulations.
        The symbol size indicates the lateral system length $L$, which is also varied randomly.
        Small symbols refer to values of $L$ closer to $5\sigma$ and big symbols to values where $L$ is closer to $20 \sigma$.
        The binodal and critical point taken from Ref.~\cite{Wilding1995} are shown in gray.
        The critical temperature $k_B T_c / \varepsilon \approx 1.188$ provides an indicator of where we might `expect' to find liquid-gas phase separation.
        Note that the systems are inhomogeneous, thus rendering the value of $\mu$ inconclusive for determining the emerging phases.
        Middle panel: an example from the training set.
        The randomly generated planar external potential $V_\mathrm{ext}(x)$ (black curve, hard walls indicated in gray) creates an inhomogeneous density profile $\rho(x)$ (blue), which is sampled in the simulation.
        The one-body direct correlation function $c_1(x)$ (orange) follows pointwise from Eq.~\eqref{eq:c1}.
        Right panel: a neural network is trained to extract and represent the underlying functional relationship $c_1(x; [\rho], T)$ in a local manner \cite{Sammueller2023neural}.
        The parametric temperature dependence and the finite-size scaling are taken into account via additional nodes in the input layer.
        We also indicate in the left panel the temperature cutoff of $k_B T / \varepsilon = 1.3$ (horizontal dotted pink line) for the case of purely supercritical training, where only simulations with higher temperature contribute---see text.
    }
    \label{fig:overview}
\end{figure*}

\subsection{Generation of training data}
\label{sec:data_generation}

Throughout this work, we consider the truncated Lennard-Jones (LJ) fluid as specified by the pairwise interaction potential
\begin{equation}
    \label{eq:LJ}
    \phi(r) = \begin{cases}
        4 \varepsilon \left[ \left(\frac{\sigma}{r}\right)^{12} - \left(\frac{\sigma}{r}\right)^{6} \right], &r \leq r_c,\\
        0, &r > r_c,
    \end{cases}
\end{equation}
for separation distance $r$.
The LJ well-depth $\varepsilon$ and particle diameter $\sigma$ set the energy and length scales, respectively.
We choose a typical truncation distance of $r_c = 2.5 \sigma$, which allows for the direct comparison of our subsequent findings to numerous simulation studies \cite{Panagiotopoulos1994,Wilding1995,Trokhymchuk1999,Errington2007,Rane2011,Evans2017}.
Note that we do not apply an energy shift in the pair potential \eqref{eq:LJ}.

Training data are acquired in grand canonical Monte Carlo (MC) simulation runs \cite{Frenkel2023} with randomly generated inhomogeneous potential energy landscapes in planar geometry, i.e.\ $V_\mathrm{ext}(\vec{r}) = V_\mathrm{ext}(x)$ \cite{Sammueller2023neural}.
We further randomize the thermodynamic state point as specified by the chemical potential $\mu$ and the temperature $T$, which are chosen uniformly within the ranges $-7 < \mu / \varepsilon < 4$ and $1 < k_B T / \varepsilon < 2$.
For ease of sampling and histogram construction, the system length in the (inhomogeneous) $x$-direction is kept fixed at a value of $L_x = 20 \sigma$.
To alleviate finite-size effects, at least to some extent, we also vary the lateral system lengths $L_y = L_z = L$ uniformly within the interval $5 < L / \sigma < 20$.
The advantages and limitations of this procedure, in particular regarding the resulting behavior in the vicinity of the critical point, are described in Sec.~\ref{sec:critical_region}.

In each simulation run, the density profile $\rho(x)$ is measured via straightforward sampling of microstates into a position-resolved histogram with bin width $\Delta x = 0.01 \sigma$.
The one-body direct correlation function then follows pointwise according to
\begin{equation}
    \label{eq:c1}
    c_1(x) = \ln\rho(x) + \beta \left[ V_\mathrm{ext}(x) - \mu \right]
\end{equation}
for all $x$ where $\rho(x) > 0$.
In total, $880$ individual simulations have been performed to gather training data for $\rho(x)$ and $c_1(x)$ profiles at different temperatures, chemical potentials, lateral system sizes, and for varying shapes of the imposed inhomogeneities in the external potential.
The total computation time of $\sim 10^4$ CPU hours for the generation of the entire data set is moderate owing to the relative ease of determining the one-body profile $\rho(x)$ in simulations.

Fig.~\ref{fig:overview} (left panel) shows the thermodynamic state points of all training simulations.
Note that we also show the coexistence curve (gray line) in the $(\mu, T)$ plane, obtained in grand canonical MC simulations by \citet{Wilding1995}.
This ends at the critical temperature $k_B T_c / \varepsilon = 1.188$ which provides an important indication of where we might hope to find phase separation using our present neural functional.
As an example of a simulation within the training set, namely at $\mu / \varepsilon \approx -0.17$ and $k_B T / \varepsilon \approx 1.76$, we show the relevant inhomogeneous one-body profiles $\rho(x)$ and $c_1(x)$ for a particular shape of $V_\mathrm{ext}(x)$.
In the right panel of Fig.~\ref{fig:overview}, we show a schematic illustration of the neural density functional mapping, which we describe in the following.

\subsection{Neural network and training procedures}
\label{sec:training}

We proceed analogously to Ref.~\cite{Sammueller2023neural} and aim at representing the direct correlation functional $c_1(x; [\rho], T)$ locally via a neural network.
That is, for a given position $x_0$, one considers the functional mapping from a section of the density profile in the vicinity of $x_0$ to the scalar value $c_1(x_0)$ of the direct correlation functional at that location, see Fig.~\ref{fig:overview} (right panel).
Hence, for each $x_0$, the density profile $\rho(x)$ is given only within a cutoff range $|x - x_0| \leq x_c$ as input to the neural network.
We deem $x_c = 3.5 \sigma$ sufficient, which leads to 701 neural input nodes for the resulting density window with the given discretization $\Delta x$ of the histograms.
The local learning of one-body direct correlations implies a quick decay of their functional dependence on the surrounding density profile, which should be valid for short-ranged interparticle interactions, provided that one stays clear of the critical region \cite{Hansen2013}.
Considering such a local functional mapping is beneficial both during training and in predictive tasks.
In particular, the neural functional remains applicable to virtually arbitrary system sizes, enabling efficient multiscale investigations \cite{Sammueller2023neural,Sammueller2024why}.
This proves to be crucial for the prediction of phase coexistence and interfacial profiles, see Secs.~\ref{sec:coexistence} and \ref{sec:eos_interface_tension}, where the system size $L_x$ must be increased substantially to yield an accurate account of liquid-gas phase separation.

To incorporate the parametric dependence of one-body direct correlation functions on temperature $T$ and lateral system size $L$, additional nodes in the input layer are provided, which accept the respective scalar values as indicated in Fig.~\ref{fig:overview}.
As a technical detail, we input $1/L$ instead of $L$, which allows us to set $1/L = 0$ during inference to extrapolate to large lateral system sizes.
We argue that the variability in $L$ has advantages over training with fixed lateral system size, although it may not account for the true finite-size scaling behavior due to insufficient information in the training data (see also Sec.~\ref{sec:critical_region} for limitations in the critical region).
Nevertheless, one may hope to avoid ingraining the specific finite-size effects of a particular choice of $L$ via this procedure.

Investigating the extrapolation capabilities of the neural network is important for practical applications but also from a conceptual point of view.
Recall that the neural functional framework relies upon extracting a functional mapping from reference data obtained for inhomogeneous equilibrium fluids, which satisfy the minimization principle \eqref{eq:omegaMinimization} by definition.
However, the underlying functional relationship might be much more general; it is not restricted a priori to true equilibrium density profiles, thereby raising profound mathematical questions about the existence of a unique continuation.
In order to scrutinize this problem from a data-driven perspective, and in particular to show how much can be learned in the absence of any possible input information about coexistence, we train a second neural functional on the basis of supercritical data only by excluding simulations where $k_B T / \varepsilon < 1.3$---see horizontal pink line in Fig.~\ref{fig:overview}.

The training routines are implemented in Keras/Tensorflow \cite{Chollet2021} following the methodology laid out in Ref.~\cite{Sammueller2023neural}.
The neural network possesses a simple mulitlayer perceptron architecture and consists of four hidden layers with 512 nodes each; we employ softplus activation functions \cite{Sammueller2023neural,Zenodo}.
Training takes approximately 30min on a recent workstation GPU.
Evaluating the trained neural functional is fast ($\sim$ ms) and can be performed on the GPU in parallel for batches of input densities and parameters.
Hence, all numerical calculations presented below are computationally inexpensive, which, e.g., facilitates mapping out whole fluid phase diagrams in seconds to minutes.
Crucially, after having trained the neural network, no further simulation results are required.
Predictions rely solely on evaluating and analysing the resulting neural representation of the one-body direct correlation functional $c_1(x; [\rho], T)$, for which we elucidate common techniques in the following.

\subsection{Neural density functional theory and functional calculus}
\label{sec:neural_theory}

From a fundamental point of view, the availability of the entire density functional relationship $c_1(\vec{r}; [\rho], T)$ suffices in principle to predict the full structural and thermodynamic behavior of any fluid model.
We lay out in this section common theoretical and numerical methods, which are taylored to the neural correlation functional and which differ in some aspects from the usual treatment of analytical (meaning explicit) free energy functionals.

The central application of classical DFT concerns the determination of the one-body inhomogeneous equilibrium density profile for a given state point $\mu$, $T$ and external potential $V_\mathrm{ext}(\vec{r})$.
One solves the Euler-Lagrange equation
\begin{equation}
    \label{eq:EL}
    \rho(\vec{r}) = \exp\left[ -\beta (V_\mathrm{ext}(\vec{r}) - \mu) + c_1(\vec{r}; [\rho], T) \right],
\end{equation}
which emerges from the minimization principle \eqref{eq:omegaMinimization} and which determines the density profile $\rho(\vec{r})$ self-consistently.
Eq.~\eqref{eq:EL} can be solved efficiently, e.g.\ with a standard mixed Picard iteration, allowing for vast parameter studies, thereby commonly outperforming many-body simulation techniques by orders of magnitude in computational cost.
The entirety of the nontrivial interparticle correlation effects are captured in the one-body direct correlation functional $c_1(\vec{r}; [\rho], T)$, which are crucial in determining the resulting equilibrium state.
Instead of approximating $c_1(\vec{r}; [\rho], T)$ analytically, see e.g.\ the mean-field functional \eqref{eq:FexcMeanField} that yields the one-body direct correlation functional explicitly upon functional differentiation \eqref{eq:c1_from_Fexc}, the machine-learning routine in Secs.~\ref{sec:data_generation} and \ref{sec:training} provides an immediate neural representation of this central functional mapping, which can be readily utilized in Eq.~\eqref{eq:EL}.

Functional differentiation of $c_1(\vec{r}; [\rho], T)$ yields information about higher-order correlations in the model fluid considered.
The two-body direct correlation functional, defined as the functional derivative
\begin{equation}
    \label{eq:c2}
    c_2(\vec{r}, \vec{r}'; [\rho], T) = \frac{\delta c_1(\vec{r}; [\rho], T)}{\delta \rho(\vec{r}')},
\end{equation}
can be evaluated efficiently from a computational representation of $c_1(\vec{r}; [\rho], T)$ with reverse mode automatic differentiation (autodiff) \cite{Baydin2018}.
This technique is particularly suited to our neural-network-based description of $c_1(\vec{r}; [\rho], T)$, as autodiff is paramount to machine learning, specifically being the central mechanism for the backpropagation of errors during training \cite{Chollet2021}.
As such, machine-learning libraries come with ready-to-use implementations that make autodiff available as an efficient high-level operation, which we leverage for the evaluation of Eq.~\eqref{eq:c2}.

Contrary to standard analytical DFT approaches, which usually commence by expressing the excess free energy $F_\mathrm{exc}([\rho], T)$ as an explicit density functional, e.g.\ in the form of the mean-field treatment \eqref{eq:FexcMeanField}, our starting point is given by the neural representation of the one-body direct correlation functional, which emerges formally as the functional derivative \eqref{eq:c1_from_Fexc}.
Nevertheless, the free energy is pertinent both in its mathematical role as a generating functional as well as for the calculation of physical quantities such as the equation of state and the surface tension (see Sec.~\ref{sec:eos_interface_tension}).
For evaluating $F_\mathrm{exc}([\rho], T)$ given a (neural) functional $c_1(\vec{r}; [\rho], T)$, we utilize functional line integration \cite{Evans1992,Sammueller2023neural,Sammueller2024why,Sammueller2024hyper}, which constitutes formally integrating the functional derivative \eqref{eq:c1_from_Fexc}.
Making the functional line integral explicit via the linear parametrization of the density profile $\rho_a(\vec{r}) = a \rho(\vec{r})$, $0 \leq a \leq 1$, gives the expression \cite{Evans1992,Sammueller2023neural}
\begin{equation}
    \label{eq:Fexc_funcintegral}
    F_\mathrm{exc}([\rho], T) = - k_B T \int\!\diff{\vec{r}} \rho(\vec{r}) \int_0^1\!\diff{a} c_1(\vec{r}; [\rho_a], T),
\end{equation}
which can be evaluated straightforwardly on the basis of the neural correlation functional.

\section{Results}
\label{sec:results}

\subsection{Bulk pair correlation functions}
\label{sec:twobody_bulk}

\begin{figure}[tbp]
    \centering
    \includegraphics[width=\columnwidth]{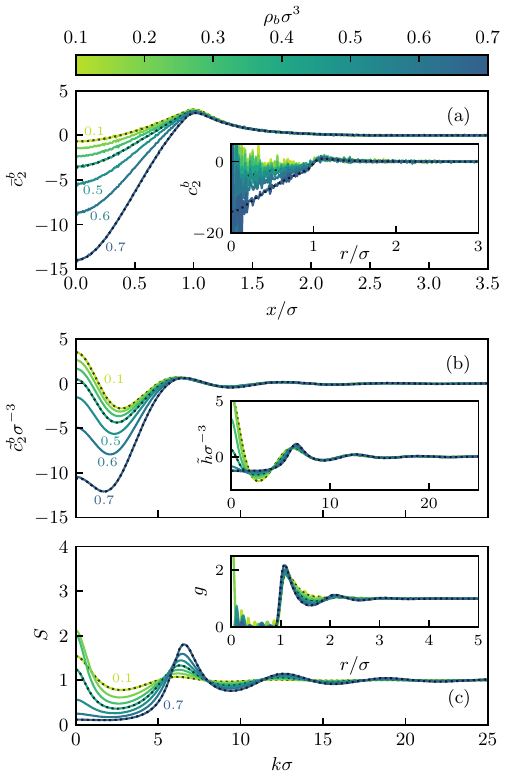}
    \caption{
        Bulk pair correlation functions from the neural functional for constant supercritical temperature $k_B T / \varepsilon = 1.5$ and different bulk densities $\rho_b \sigma^3 = 0.1, 0.2, 0.3, 0.4, 0.5, 0.6$, and $0.7$, as indicated by ticks on the colorscale and labeling of the curves.
        Shown are (a) the planar two-body direct correlation function $\bar{c}_2^b(x)$ along with the radial representation $c_2^b(r)$ (inset), (b) its Fourier transform $\tilde{c}_2^b(k)$ and the total correlation function $\tilde{h}(k)$ in Fourier space (inset) as obtained from the OZ equation \eqref{eq:OZ}, (c) the static structure factor $S(k)$ and the radial distribution function $g(r)$ (inset).
        For three bulk densities $\rho_b \sigma^3 = 0.1, 0.4$, and $0.7$, we show results for each quantity (dotted lines), obtained from Fourier transform and OZ inversion of bulk grand canonical MC simulation data for $g(r)$.
    }
    \label{fig:twobody_bulk}
\end{figure}

As a first investigation, we deliberately stay clear of any liquid-gas phase transition and consider pair correlations in bulk at constant supercritical temperature $k_B T / \varepsilon = 1.5$.
Evaluating Eq.~\eqref{eq:c2} via autodiff with constant density input $\rho(\vec{r}) = \rho_b$ and exploiting translational invariance yields the bulk two-body direct correlation function $\bar{c}_2^b(x)$ in \emph{planar} geometry (indicated here and in the following by the overbar); we drop the parametric dependence on temperature in the notation.

The numerical result can be transformed to the standard radial representation $c_2^b(r)$ by writing out the lateral integration
\begin{equation}
    \label{eq:c2_radial_to_planar}
    \begin{split}
        \bar{c}_2^b(x) &= \int\!\diff{y}\!\int\!\diff{z} c_2^b\left(r = \sqrt{x^2 + y^2 + z^2}\right)\\
            &= 2 \pi \int_x^\infty\!\diff{r} r c_2^b(r)
    \end{split}
\end{equation}
that arises due to the planar geometrical setup.
Differentiation of Eq.~\eqref{eq:c2_radial_to_planar} with respect to $x$ gives the inverse transformation
\begin{equation}
    \label{eq:c2_planar_to_radial}
    c_2^b(r) = - \frac{1}{2 \pi r} \left.\frac{\diff{\bar{c}_2^b(x)}}{\diff x}\right|_{x = r}.
\end{equation}

The total correlation function is obtained via the Ornstein-Zernike (OZ) route.
A one-dimensional Fourier transform of $\bar{c}_2^b(x)$ yields the radial quantity $\tilde{c}_2^b(k)$ in Fourier space (indicated by the tilde).
The Ornstein-Zernike equation
\begin{equation}
    \label{eq:OZ}
    \tilde{h}(k) = \frac{\tilde{c}_2^b(k)}{1 - \rho_b \tilde{c}_2^b(k)}
\end{equation}
then determines the Fourier transform of the total correlation function $h(r)$ algebraically, from which the static structure factor follows as
\begin{equation}
    \label{eq:S}
    S(k) = 1 + \rho_b \tilde{h}(k).
\end{equation}
The radial distribution function $g(r) = h(r) + 1$ is determined by a radial Fourier(-Hankel) backtransform of $\tilde{h}(k)$ to real space:
\begin{equation}
    \label{eq:Fourier-Hankel_back}
    h(r) = \frac{1}{2 \pi^2 r} \int_0^\infty\!\diff{k} k \sin(kr) \tilde{h}(k).
\end{equation}

Results for the various bulk pair correlation functions are shown in Fig.~\ref{fig:twobody_bulk} for a supercritical temperature of $k_B T / \varepsilon = 1.5$ and different bulk densities.
See also Ref.~\cite{Sammueller2024pair} for comparative data obtained with isothermal training and from other methods, e.g.\ pair-correlation matching \cite{Dijkman2024}, and Ref.~\cite{Sammueller2023neural} for pair correlations of the hard sphere fluid.
From Fig.~\ref{fig:twobody_bulk} we note: i) noisy artifacts arise in $c_2^b(r)$ and $g(r)$ for $r < \sigma$ due to numerical intricacies associated with the transformations \eqref{eq:c2_planar_to_radial} and \eqref{eq:Fourier-Hankel_back}, ii) $c_2^b(r)$ and $\bar{c}_2^b(x)$ are independent of density $\rho_b$ for $r > \sigma$, reflecting the fact that $c_2^b(r)$ quickly reaches its asymptotic limit $- \beta \phi(r)$ \cite{Dijkstra2000}, iii) $S(k)$ exceeds $2.0$ for small wave numbers $k$ at reduced densities $\rho_b \sigma^3 = 0.2, 0.3$, a possible sign of the approach to a critical point, and iv) the neural predictions match very closely the simulation results extracted from separate grand canonical MC for $g(r)$---see dotted lines referring to three specific densities in Fig.~\ref{fig:twobody_bulk}.
We emphasize that \emph{no} information regarding pair correlations was incorporated during training.

\subsection{Lines of maximal isothermal compressibility and correlation length, Fisher-Widom line and spinodal}
\label{sec:map_T_rho}

\begin{figure*}[tbp]
    \centering
    \includegraphics{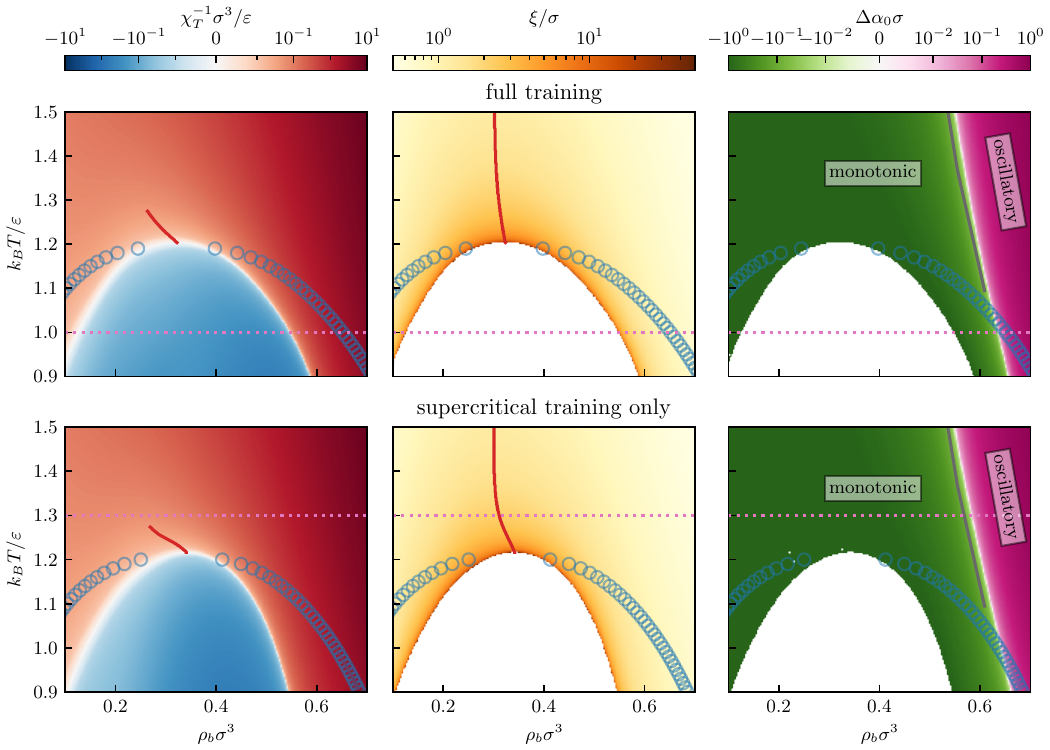}
    \caption{
        First column: the inverse of the isothermal compressibility obtained from $\tilde{c}_2^b(k = 0)$ according to Eq.~\eqref{eq:chiinv_from_c2}.
        The spinodal corresponds to $\chi_T^{-1} = 0$, and the predicted values of $\chi_T^{-1}$ inside of it are negative.
        The line of maximal isothermal compressibility is indicated in red.
        Second column: true correlation length $\xi$ from pole analysis according to Eq.~\eqref{eq:xi_from_pole}.
        The Widom line of largest correlation length is indicated in red.
        Third column: difference $\Delta \alpha_0$ between the imaginary parts of the leading monotonic and oscillatory poles, which determines the Fisher-Widom line via $\Delta \alpha_0 = 0$.
        On the low density side (green) the ultimate decay of $h(r)$ is monotonic whereas on the high density side it is damped oscillatory (purple).
        Simulation results from Ref.~\cite{Dijkstra2000} are reproduced (black line) after applying corresponding states rescaling, see text.
        The blue circles denote neural liquid-gas coexistence densities, see Fig.~\ref{fig:binodal_comparison} below.
        Results obtained from a neural correlation functional that has been trained with data including subcritical temperatures (first row) as well as from purely supercritical training (second row) are shown.
        The lowest temperatures of the simulations that contribute to the training are indicated by the horizontal dotted pink lines.
    }
    \label{fig:map_T_rho}
\end{figure*}

In bulk, the inverse of the isothermal compressibility $\chi_T$ is accessible from the pair direct correlation function:
\begin{equation}
    \label{eq:chiinv_from_c2}
    \chi_T^{-1} = k_B T \rho_b (1 - \rho_b \tilde{c}_2^b(k = 0)).
\end{equation}

Although we obtain the planar quantity $\bar{c}_2^b(x)$ from the autodifferentiated neural functional via Eq.~\eqref{eq:c2}, we have shown that a simple one-dimensional Fourier transform suffices to acquire $\tilde{c}_2^b(k)$ in \emph{radial} geometry \cite{Sammueller2023neural}.
Hence, evaluation at $k = 0$ gives direct access to $\chi_T^{-1}$ and results are shown in Fig.~\ref{fig:map_T_rho} in the temperature-density plane.
Strikingly, the neural functional predicts a spinodal where $\chi_T^{-1} = 0$ and which bounds an unstable region where $\chi_T < 0$.
The first column in Fig.~\ref{fig:map_T_rho} shows the (red) line of maximal compressibility obtained from scans of varying density at fixed temperature.
The quantity $\chi_T^{-1}$ vanishes as $T$ is reduced and below a critical value the spinodal emerges.
In the first row results are presented for training above $k_B T / \varepsilon = 1.0$ whereas the second row includes only training data at supercritical temperatures, $k_B T / \varepsilon > 1.3$.
Note that both training protocols deliver a spinodal, albeit with slightly different critical points.

We turn now to the asymptotic (large $r$) decay of $h(r)$.
Specifically, we determine both the Widom line of maximal true correlation length, middle column, and the Fisher-Widom crossover line, third column of Fig.~\ref{fig:map_T_rho}.
A pole analysis of the total correlation function determines the asymptotic behavior of $h(r)$ at long range, $r \rightarrow \infty$, where either (damped) oscillatory or monotonic decay is exhibited \cite{Dijkstra2000, Fisher1969, Evans1993}.
In Fourier space, the poles $\alpha = \alpha_1 + i \alpha_0$ of the total pair correlation function are readily determined by the OZ equation \eqref{eq:OZ}, leading back to the analysis of the direct correlation function $\tilde{c}_2^b(k)$.
The zeroes of the denominator $1 - \rho_b \tilde{c}_2^b(\alpha)$ in Eq.~\eqref{eq:OZ} yield the conditions \cite{Dijkstra2000}
\begin{align}
    \label{eq:poles1}
    1 &= 4 \pi \rho_b \int_0^\infty\!\diff{r} r^2 c_2^b(r) \frac{\sinh(\alpha_0 r)}{\alpha_0 r} \cos(\alpha_1 r),\\
    \label{eq:poles2}
    1 &= 4 \pi \rho_b \int_0^\infty\!\diff{r} r^2 c_2^b(r) \cosh(\alpha_0 r) \frac{\sin(\alpha_1 r)}{\alpha_1 r},
\end{align}
where $c_2^b(r)$ is the bulk two-body direct correlation function in direct space and in \emph{radial} geometry.
A pure imaginary pole gives rise to monotonic decay and is given by Eq.~\eqref{eq:poles1} with $\alpha_1 = 0$; Eq.~\eqref{eq:poles2} does not arise in this case.

It remains to express Eqs.~\eqref{eq:poles1} and \eqref{eq:poles2} in terms of $\bar{c}_2^b(x)$ instead of $c_2^b(r)$ in order to accommodate the planar geometrical setup.
The identity \eqref{eq:c2_planar_to_radial} enables us to rewrite Eqs.~\eqref{eq:poles1} and \eqref{eq:poles2}, via partial integration, as
\begin{align}
    \label{eq:poles1_planar}
    \begin{split}
        1 &= 2 \rho_b \int_0^\infty\!\diff{r} \bar{c}_2^b(x = r) \Bigl[ \cosh(\alpha_0 r) \cos(\alpha_1 r)\\
        &\qquad \qquad - \frac{\alpha_1}{\alpha_0} \sinh(\alpha_0 r) \sin(\alpha_1 r) \Bigl],
    \end{split}\\
    \label{eq:poles2_planar}
    \begin{split}
        1 &= 2 \rho_b \int_0^\infty\!\diff{r} \bar{c}_2^b(x = r) \Bigl[ \frac{\alpha_0}{\alpha_1} \sinh(\alpha_0 r) \sin(\alpha_1 r)\\
        &\qquad \qquad + \cosh(\alpha_0 r) \cos(\alpha_1 r) \Bigl].
    \end{split}
\end{align}

A pure imaginary (monotonic) pole $a_0^\mathrm{mon}$ is determined by the simpler condition
\begin{equation}
    \label{eq:first_pole_planar}
    1 = 2 \rho_b \int_0^\infty\!\diff{r} \bar{c}_2^b(x = r) \cosh(\alpha_0^\mathrm{mon} r),
\end{equation}
which follows from setting $\alpha_1 = 0$ either directly in Eq.~\eqref{eq:poles1_planar} or, consistently, in Eq.~\eqref{eq:poles1} and performing the partial integration using Eq.~\eqref{eq:c2_planar_to_radial}.
Solving Eq.~\eqref{eq:first_pole_planar} along with the coupled Eqs.~\eqref{eq:poles1_planar} and \eqref{eq:poles2_planar} for a complex (oscillatory) pole is implemented efficiently in terms of a minimization procedure.

At low bulk densities, and in the neighbourhood of the critical point, one expects monotonic asymptotic decay of the total pair correlation function: $h(r) \sim \exp(-\alpha_0^\textrm{mon} r) / r$ at large $r$.
It follows that the leading pole, i.e.\ with the smallest imaginary part, will be given by Eq.~\eqref{eq:first_pole_planar}.
The true correlation length is
\begin{equation}
    \label{eq:xi_from_pole}
    \xi = \frac{1}{\alpha_0^\mathrm{mon}},
\end{equation}
which is plotted for a range of bulk densities and temperatures in the second column in Fig.~\ref{fig:map_T_rho}.
The behavior of $\xi$ is consistent with the results for the isothermal compressibility $\chi_T$: the correlation length increases rapidly when approaching the critical region and indicates a spinodal where $\xi$ diverges when continuing to lower the temperature.
Tracing $\xi \rightarrow \infty$ or $\chi_T \rightarrow \infty$ yields numerically identical spinodals.
Of course, this is expected since setting $\alpha_0^\mathrm{mon}$ to zero in Eq.~\eqref{eq:first_pole_planar} is equivalent to requiring the right hand side of Eq.~\eqref{eq:chiinv_from_c2} to vanish.
Within the spinodal, Eq.~\eqref{eq:first_pole_planar} yields no solution and $\xi$ is undefined.

The difference between the imaginary parts of the leading monotonic and oscillatory ($\tilde{\alpha}$) poles determines whether monotonic or oscillatory decay pertains at longest range, which in turn determines the Fisher-Widom line via the condition $\Delta \alpha_0 = \alpha_0^\mathrm{mon} - \tilde{\alpha}_0 = 0$, see Fig.~\ref{fig:map_T_rho}, third column.
The black line in the third column denotes the MC simulation results of \citet{Dijkstra2000} obtained for a truncated and shifted LJ potential rescaled to take account of how the critical point alters when shifting the potential at cutoff.
Note how close their Fisher-Widom line lies to our present neural functional prediction (see white line where $\Delta \alpha_0 = 0$) and that determining the Fisher-Widom line is a subtle task requiring accurate knowledge of the decay of bulk pair correlation functions not easily accessible from simulations.
That our neural functional discovers this crossover line, never having encountered directly bulk pair correlation functions in the learning process, is remarkable.
The Fisher-Widom line is estimated to cross the liquid branch of the neural coexistence curve, described in the next subsection and denoted by blue circles in Fig.~\ref{fig:map_T_rho}, at a temperature of $k_B T / \varepsilon \approx 1.04$.
This has implications for the structure of the liquid-gas interface, see Sec.~\ref{sec:eos_interface_tension}.

\subsection{Liquid-gas coexistence, binodal and estimate of critical point}
\label{sec:coexistence}

\begin{figure*}[tb]
    \centering
    \includegraphics{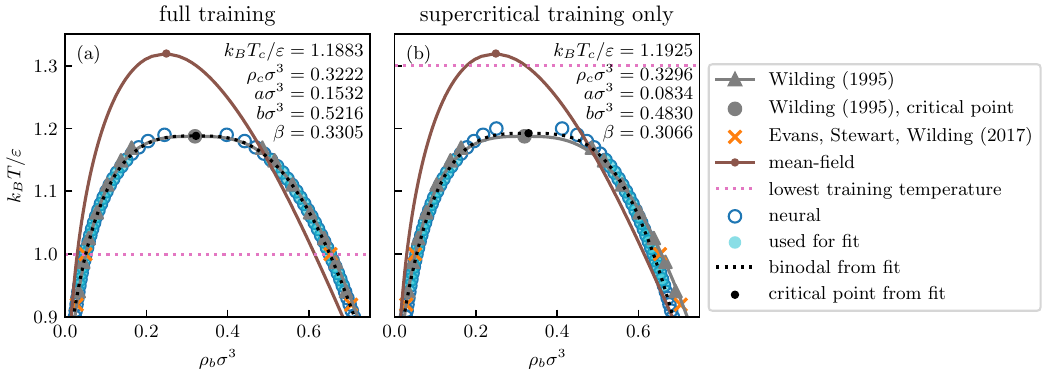}
    \caption{
        Liquid-gas coexistence densities (blue circles) from neural minimization of the free interface, obtained with neural correlation functionals trained (a) on the basis of the whole reference data set, $k_B T / \varepsilon > 1.0$, and (b) with supercritical data only, $k_B T / \varepsilon > 1.3$.
        The temperature cutoffs are indicated by the pink horizontal lines.
        For comparison, we show data from Refs.~\cite{Wilding1995,Evans2017} as well as the binodal from the standard analytical DFT treatment \cite{Evans2017} of the LJ fluid based on the Rosenfeld \cite{Rosenfeld1989} hard-sphere FMT functional plus mean-field attraction (brown line).
        Fitting the neural coexistence data in panel (a) to Eq.~\eqref{eq:binodal} yields the dotted black line (binodal) with the parameters displayed.
        The resulting estimates of the critical point and the binodal are close to those of \citet{Wilding1995} (gray).
        Panel (b) shows corresponding results for the case of purely supercritical training.
    }
    \label{fig:binodal_comparison}
\end{figure*}

\begin{figure}[tb]
    \centering
    \includegraphics{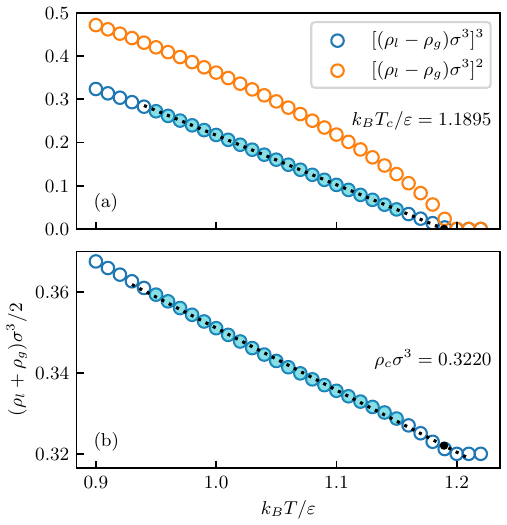}
    \caption{
        Critical temperature and density (black dots) obtained by extrapolation of (a) the cubed difference of liquid-gas densities $\rho_l$ and $\rho_g$, and of (b) the rectilinear diameter law.
        Data points utilized for the linear regressions are indicated in cyan and the dotted black lines depict the resulting fit functions.
        The results for the critical point and the binodal parameters are consistent with Fig.~\ref{fig:binodal_comparison}.
        We also show in (a) the squared difference of the the coexisting densities (the mean-field prediction)---see text.
    }
    \label{fig:binodal_scaling}
\end{figure}

In order to investigate liquid-gas phase coexistence, the neural direct correlation functional is used in the iteration of the Euler-Lagrange equation \eqref{eq:EL} keeping the mean density $\bar{\rho} = \int \diff{x} \rho(x) / L_x$ fixed.
When initialized with a step-like density profile at sufficiently low temperature, the minimization yields a phase-separated system with a liquid and a gas domain, from which one can determine the coexisting densities provided that the system has been chosen large enough.
We set $L_x = 100 \sigma$ for the following investigations and hence exploit here the multiscale applicability of the neural functional.
Performing this procedure at different temperatures allows us to trace the binodal, which is shown and compared with simulation data in Fig.~\ref{fig:binodal_comparison}.

One can attempt to fit the binodal and determine the critical point in various ways that can incorporate critical exponents beyond mean-field.
Following \citet{Wilding1995}, we exclude the near critical region and take the neural coexistence densities within the temperature range $0.95 \leq k_B T / \varepsilon \leq 1.15$ to fit the binodal via \cite{Wilding1995}
\begin{equation}
    \label{eq:binodal}
    \rho_\pm = a |T^* - T^*_c| \pm b |T^* - T^*_c|^\beta + \rho_c
\end{equation}
with scaled temperature $T^* = k_B T / \varepsilon$, liquid/gas densities $\rho_+ = \rho_l$ and $\rho_- = \rho_g$, and critical density $\rho_c$, temperature $T_c$, exponent $\beta$, and amplitudes $a$ and $b$.
Note that the exponent $\beta$ should not be confused with the inverse temperature.
Eq.~\eqref{eq:binodal} is empirical; it fails at low temperatures.

Unlike in Ref.~\cite{Wilding1995}, the critical point is, a priori, undetermined in Eq.~\eqref{eq:binodal}; $\rho_c$ and $T_c^*$ are to be deduced in the fit along with all other parameters.
To demonstrate the robustness of the fitting procedure, we also keep the (critical) exponent $\beta$ as a free parameter, although we bear in mind the Ising result $\beta = 0.32630(22)$ \cite{Ferrenberg2018}.
The results of the binodal fit are shown in Fig.~\ref{fig:binodal_comparison} and agree very well with the data from highly accurate simulations \cite{Wilding1995}.
This procedure, which focuses on coexistence densities at temperatures sufficiently far below the critical point such that the correlation length $\xi$ is less than the system size, attempts to avoid some of the intricacies of the critical region that are also pertinent for the neural functional (see Sec.~\ref{sec:critical_region}).
We note that the resulting value of the `critical' exponent $\beta$ is, from full training, 0.330 which lies close to the Ising value \cite{Ferrenberg2018}, demonstrating that the fit is meaningful as well as indicating \emph{possible} beyond-mean-field character of the neural functional.
Note also that the corresponding parameters/results from Wilding's MC simulations and fitting \cite{Wilding1995}: $k_B T_c / \varepsilon = 1.188$, $\rho_c \sigma^3 = 0.320$, $a \sigma^3 = 0.182$, $b \sigma^3 = 0.523$, $\beta = 0.326$, are close to the present.

Following the work of \citet{Panagiotopoulos1994}, an alternative estimate of the critical temperature $T_c$ and density $\rho_c$ proceeds by a regression and extrapolation of the cubed difference of coexistence densities and of the rectilinear diameter law, respectively.
As before, coexistence data in the close vicinity of the critical point needs to be excluded.
The thrust of this approach is the prediction that $\rho_l - \rho_g$ should decay approximately as $(T_c - T)^{1/3}$ near the critical point rather than as $(T_c - T)^{1/2}$, which is the mean-field prediction.
Fig.~\ref{fig:binodal_scaling}(a) indicates that our neural functional yields results consistent with non-mean-field behavior and the estimate of the critical temperature is close to the simulation result.
Fig.~\ref{fig:binodal_scaling}(b) plots the rectilinear diameter versus temperature which allows for an estimate of the critical density that is again close to the simulation result.
The results shown in Fig.~\ref{fig:binodal_scaling} are consistent with the previous fit to the binodal using Eq.~\eqref{eq:binodal}.

\subsection{Bulk equation of state, liquid-gas interface and surface tension}
\label{sec:eos_interface_tension}

\begin{figure}[tb]
    \centering
    \includegraphics{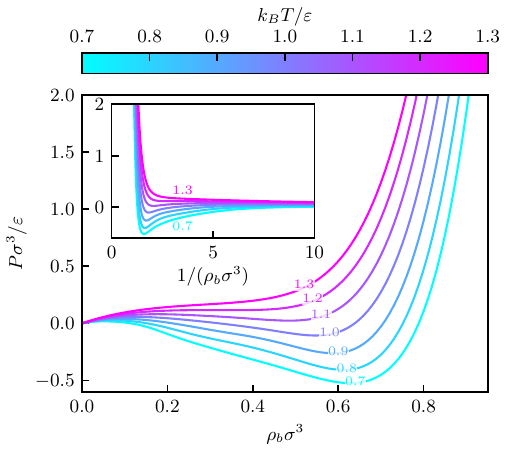}
    \caption{
        Bulk equation of state: pressure $P$ as a function of the bulk density $\rho_b$ for temperatures $k_B T / \varepsilon = 0.7, 0.8, 0.9, 1.0, 1.1, 1.2$, and $1.3$, as labeled and indicated on the colorscale.
        The inset shows the same data plotted with respect to the inverse bulk density, i.e.\ proportional to the system volume.
        Note the appearance of a van der Waals loop at subcritical temperatures.
    }
    \label{fig:eos}
\end{figure}

\begin{figure}[tb!]
    \centering
    \includegraphics{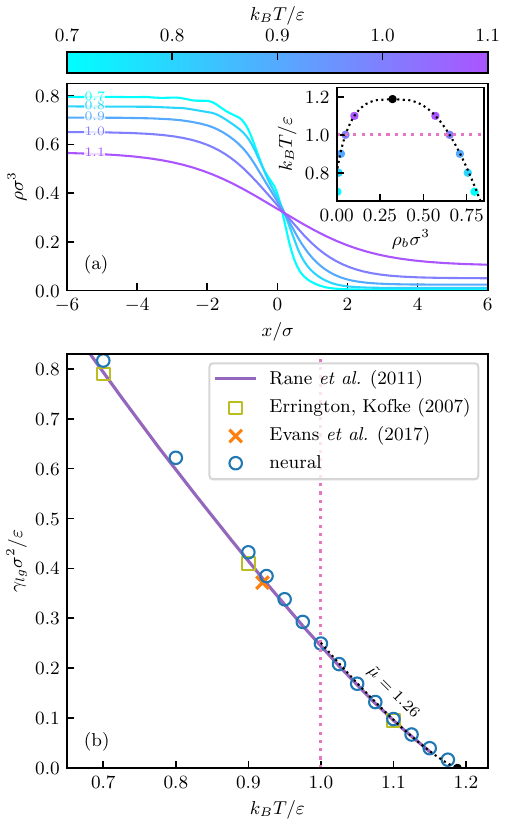}
    \caption{
        (a) Liquid-gas interfacial density profiles as obtained from solution of Eq.~\eqref{eq:EL} for a phase-separated system upon keeping the mean density fixed.
        Results are shown for five temperatures $k_B T / \varepsilon = 0.7, 0.8, 0.9, 1.0$, and $1.1$, see labels.
        The inset indicates the respective bulk coexistence densities, the fitted binodal (dotted black line) and critical point (black dot) reproduced from Fig.~\ref{fig:binodal_comparison}; the horizontal dotted pink line shows the temperature cutoff of $k_B T / \varepsilon = 1.0$ during training.
        Oscillations of the interfacial density profile that extend into the liquid domain are visible for the two lowest temperatures.
        (b) Surface tension $\gamma_{lg}$ of the liquid-gas interface as determined from functional line integration of interfacial density profiles.
        Simulation results are taken from Refs.~\cite{Rane2011,Errington2007,Evans2017}.
        The cutoff for training, $k_B T / \varepsilon = 1.0$, is indicated again, now by the vertical dotted pink line.
        Fitting to neural data yields the effective exponent for the tension $\tilde\mu \approx 1.26$---see text.
    }
    \label{fig:tension}
\end{figure}

Evaluating the excess free energy on the basis of the neural direct correlation functional is facilitated via functional line integration, for which Eq.~\eqref{eq:Fexc_funcintegral} provides a straightforward parametrization.
We utilize the efficient access to such thermodynamic information for both homogeneous (bulk) and inhomogeneous systems in the following to obtain the bulk equation of state and the surface tension of the liquid-gas interface.

In bulk, evaluation of Eq.~\eqref{eq:Fexc_funcintegral} for constant target density $\rho(\vec{r}) = \rho_b$ yields the excess free energy density $\psi_b = F_\mathrm{exc}([\rho_b], T) / V$, where $V$ is the system volume.
This allows us to calculate the bulk pressure
\begin{equation}
    \label{eq:eos}
    P(\rho_b, T) = k_B T \rho_b (1 - c_1^b) - \psi_b,
\end{equation}
where $c_1^b = c_1(x; [\rho_b], T)$ follows from direct evaluation of the neural correlation functional with constant density input and at arbitrary position $x$ due to translational invariance.
Note that the excess chemical potential can be identified as $\mu_\mathrm{exc} = - k_B T c_1^b$.
We show the neural prediction for the equation of state $P(\rho_b, T)$, obtained from Eq.~\eqref{eq:eos}, for a range of temperatures in Fig.~\ref{fig:eos}.
Note that a van der Waals loop emerges for subcritical temperatures, i.e.\ for $k_B T / \varepsilon \lesssim 1.2$ and that negative pressures arise at low temperatures.
The inset to Fig.~\ref{fig:eos} plots pressure versus volume.
Performing a Maxwell equal area construction yields, within the expected numerical accuracy, values of the coexisting liquid and gas densities equal to those determined in Sec.~\ref{sec:coexistence} by direct solution of the Euler-Lagrange equation for an inhomogeneous phase-separated system.
This attests to the consistency of the neural functional.

As the functional line integral \eqref{eq:Fexc_funcintegral} is constructed for inhomogeneous density input, evaluation with the liquid-gas interfacial density profiles provides direct access to the surface tension.
Determining $F_\mathrm{exc}([\rho], T)$ from Eq.~\eqref{eq:Fexc_funcintegral} facilitates to calculate the grand potential $\Omega([\rho], T)$ via Eq.~\eqref{eq:omegaFunctional}, which in turn yields the liquid-gas tension as the surface excess grand potential per unit area:
\begin{equation}
    \label{eq:tension}
    \gamma_{lg} = \frac{\Omega + P V}{A}
\end{equation}
with system volume $V$, lateral system area $A$ and pressure $P$ at coexistence as given by Eq.~\eqref{eq:eos}.
We note that subtle discrepancies in the values of the liquid and gas pressures $P_l$ and $P_g$ arise for low temperatures due to the accumulation of numerical errors when integrating through the van der Waals loop in Eq.~\eqref{eq:eos}.
To alleviate this numerical issue, we compute $P V = P_l V_l + P_g V_g$ and employ this relation consistently in Eq.~\eqref{eq:tension}; $V_l$ and $V_g$ denote the volumes of liquid and gas domains, respectively.
This procedure is inline with the definition of the surface tension $\gamma_{lg}$ as a genuine excess quantity.
We also note that the pressures predicted at bulk coexistence are generally small, with values of the order of $10^{-2} \varepsilon / \sigma^3$.

Neural predictions of the interfacial density profile and the surface tension for a range of temperatures are shown in Fig.~\ref{fig:tension}.
The density profiles shown in panel (a) exhibit damped oscillatory decay into the bulk liquid for the two lowest temperatures plotted: $k_B T / \varepsilon = 0.7$ and $0.8$.
Both correspond to temperatures well below that where the Fisher-Widom line meets the liquid binodal, i.e.\ $k_B T / \varepsilon \approx 1.04$, see Sec.~\ref{sec:map_T_rho}.
For higher temperatures the decay of the density profile into bulk appears to be monotonic.
Such a scenario is similar to an early DFT study by \citet{Evans1993} for a square-well model fluid.
We return to the issue of how oscillations in the density profile might be eroded by capillary wave fluctuations in the discussion, cf.\ Sec.~\ref{sec:discussion_physics}.
The values obtained for the surface tension are close to state of the art simulation data \cite{Rane2011,Errington2007,Evans2017}, notably also for temperatures significantly below the training cutoff $k_B T / \varepsilon = 1.0$---see the vertical pink dotted line in Fig.~\ref{fig:tension}.
This demonstrates both the validity of the minimization \eqref{eq:EL} for obtaining liquid-gas density profiles as well as of the functional integration \eqref{eq:Fexc_funcintegral} for the evaluation of $\gamma_{lg}$.
Both procedures remain robust when extrapolating to lower temperatures, see Sec.~\ref{sec:supercritical} for further discussion.

The data in Fig.~\ref{fig:tension} obtained for a range of subcritical temperatures allows us to investigate the near-critical behavior of the surface tension.
Recall that the latter should vanish as $\gamma_{lg} \sim (T_c - T)^{\tilde\mu}$ with a critical exponent $\tilde\mu$, given by $\tilde\mu = 2 \nu \approx 1.26$ for the three-dimensional Ising case, where the correlation length diverges as $\xi \sim (T_c - T)^{-\nu}$.
We proceed analogously to Sec.~\ref{sec:coexistence} and perform a fit of subcritical data within the same temperature range $0.95 \leq k_B T / \varepsilon \leq 1.15$.
Importantly, we thereby fix the critical temperature $k_B T_c / \varepsilon = 1.1883$ as obtained from the binodal regression in Fig.~\ref{fig:binodal_comparison}.
This choice of fitting procedure yields the `effective' critical exponent $\tilde{\mu} \approx 1.26$, consistent with the Ising value.
Recall that the mean-field exponent is $\tilde{\mu} = 3/2$.
We also examined the `10-90' width of the liquid-gas interface in the same temperature range.
This appears to diverge in the same fashion as the correlation length, i.e.\ with exponent $\nu \approx 0.63$, as expected from scaling arguments.

We conclude that using density profiles of the free interface, in a suitably chosen range of subcritical temperatures, allows one to perform fits to both the coexistence densities, cf.\ Fig.~\ref{fig:binodal_comparison}, and the surface tension, cf.\ Fig.~\ref{fig:tension}, that suggest non-mean-field behavior.
Of course, in making such fits one is deliberately avoiding the direct evaluation of the neural functional for temperatures very close to $T_c$, where the correlation length becomes very long.
We might expect results to cross over to mean-field behavior, as a result of the finite size of our systems as we lay out in the following section.

\subsection{Critical region and Ornstein-Zernike plots}
\label{sec:critical_region}

\begin{figure}[tbp]
    \centering
    \includegraphics[width=\linewidth]{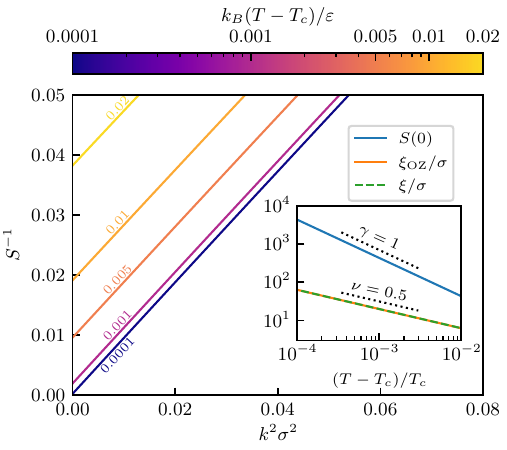}
    \caption{
        OZ plot of the inverse structure factor $S$ with respect to $k^2$ for supercritical temperatures at constant critical density $\rho_c$, from which one can determine $S(0)$, the short-range correlation length $R$, and the OZ correlation length $\xi_\mathrm{OZ}$---see text.
        Plots of $S(0)$ and $\xi_\mathrm{OZ}$ with respect to $|T-T_c|/T_c$ determine the critical exponents $\gamma$ and $\nu$ (inset); our neural functional yields the mean-field values as shown.
        The true correlation length $\xi$, calculated from Eq.~\eqref{eq:xi_from_pole}, lies very close to $\xi_\mathrm{OZ}$ for temperatures close to $T_c$, as shown (in green) in the inset.
    }
    \label{fig:S_OZ}
\end{figure}

Although the neural functional is straightforward to evaluate for state points in the vicinity of the critical point, the resulting predictions must be treated with particular caution.
The limitations are already apparent by comparing results obtained for the spinodal, see the isothermal compressibility $\chi_T$ and correlation length $\xi$ in Fig.~\ref{fig:map_T_rho}, and the estimate of the critical point from fitting to the binodal using Eq.~\eqref{eq:binodal} in Fig.~\ref{fig:binodal_comparison}.
One observes that the critical point obtained from the latter does not match exactly the top of the spinodal from the former.
To be precise, we get $k_B T_c / \varepsilon \approx 1.188$ from the binodal fit in Fig.~\ref{fig:binodal_comparison} but $k_B T_c / \varepsilon \approx 1.203$ from the locus of either $\chi_T^{-1} = 0$ or $\xi^{-1} = 0$ in Fig.~\ref{fig:map_T_rho}.
It is important to note that simply tracing the neural binodal (blue circles) in Fig.~\ref{fig:binodal_comparison} will also yield the identical critical temperature $k_B T_c / \varepsilon \approx 1.203$ and critical density as obtained from the spinodal, as is expected from a DFT approach.
On the other hand, our determination of the critical temperature, $k_B T_c / \varepsilon \approx 1.188$, obtained by fitting to a suitably chosen range of subcritical data, is inspired by simulation studies \cite{Panagiotopoulos1994,Wilding1995} and deliberately invokes additional physics by excluding results in the near-critical region that will be beset by finite-size artifacts.

We first should place our results in perspective: we are focusing on estimates of the critical point, obtained via our neural functional that had no prior knowledge of the existence of phase separation let alone a critical point, which lie within about one percent of the most accurate, independent, simulation study of the critical point of the LJ fluid \cite{Wilding1995} which finds $k_B T_c / \varepsilon = 1.188$.

In the following, we scrutinize further issues that arise when inferring properties in the critical region.
Recall that, for a suitably chosen temperature range, the neural functional yields coexisting densities that are fitted better, within our procedures, by a non mean-field critical exponent close to the Ising value $\beta \approx 0.33$ than by the mean-field result $\beta = 1/2$, see Sec.~\ref{sec:coexistence}.
Similarly, for the surface tension we find via an analogous procedure the critical exponent $\tilde{\mu} \approx 1.26$ instead of the mean-field prediction $\tilde{\mu} = 3/2$.
Of course, when approaching $T_c$ very closely the physics is more subtle.
For example, \citet{Panagiotopoulos1994} argues that certain properties in the critical region could be forced to appear mean-field-like because of the finite size of the simulation.
The basic idea is that one cannot access correlation lengths that are larger than or of the same order as the simulation box size.
In our study we provide training data from MC simulations of planar density profiles in boxes of lateral size $L < 20 \sigma$.
It is clear that large correlation lengths will be suppressed, and one might therefore expect the neural functional to inherit mean-field critical behavior.

We choose to examine this issue by performing an OZ plot of the static structure factor, cf.\ Sec.~\ref{sec:twobody_bulk} and Eq.~\eqref{eq:S}, in the super-critical regime, employing the traditional OZ description $S(k) = S(0) / (1 + \xi_\mathrm{OZ}^2 k^2)$ for small wavenumbers $k$.
The inverse of $S(k)$ is plotted against $k^2$, at supercritical temperatures approaching $T_c$, and at fixed critical density $\rho_c \sigma^3 = 0.322$, in Fig.~\ref{fig:S_OZ}.
Note that the reference value of the critical temperature refers here to $k_B T_c / \varepsilon \approx 1.2031$ as given by the maximum of the spinodal, defined, of course, by $1/S(0) = 0$.

From the OZ plot, we can investigate the isothermal compressibility critical exponent $\gamma$ by analysing the scaling of the $y$-intercept $1 / S(0)$, which determines $S(0) \sim \chi_T \sim |T-T_c|^{-\gamma}$.
The slope of the lines in Fig.~\ref{fig:S_OZ} determines the OZ correlation length $\xi_\mathrm{OZ}$ given by
\begin{equation}
    \label{eq:corr_OZ}
    \xi_\mathrm{OZ}^2 = R^2 S(0),
\end{equation}
where the short-range correlation length $R$ is the second moment of $c_2^b(r)$:
\begin{equation}
    \label{eq:R_OZ}
    \begin{split}
        R^2 &= \frac{2 \pi \rho_b}{3} \int_0^\infty \!\diff{r} r^4 c_2^b(r)\\
        &= \rho_b \int_0^\infty \!\diff{x} x^2 \bar{c}_2^b(x),
    \end{split}
\end{equation}
and the second equation, involving the planar two-body direct correlation function, follows via Eq.~\eqref{eq:c2_planar_to_radial} and partial integration.
The gradient of $1 / S(k)$ in the OZ plot is $R^2$.
From Fig.~\ref{fig:S_OZ} we observe this is constant close to the critical temperature; we find $R / \sigma \approx 0.97$.

Defining the critical exponent $\nu$ by $\xi_\mathrm{OZ} \sim |T-T_c|^{-\nu}$ it follows that our analysis predicts $\gamma = 2 \nu$, since $R$ remains finite.
This implies a (Fisher) exponent $\eta = 0$.
The inset in Fig.~\ref{fig:S_OZ} shows a log-log plot of $S(0)$ and the OZ correlation length $\xi_\mathrm{OZ}$ as functions of the reduced supercritical temperature difference $(T - T_c) / T_c$.
Our neural functional results yield the mean-field critical exponents $\gamma = 1$ and $\nu = 0.5$ to high accuracy, in keeping with ideas of \citet{Panagiotopoulos1994} and others.

We also investigated the critical behavior of the true correlation length $\xi$ as given by the pole analysis, see Sec.~\ref{sec:map_T_rho} and Eq.~\eqref{eq:xi_from_pole}.
As expected, see the inset, $\xi$ and $\xi_\mathrm{OZ}$ are almost identical near the critical point.
Hence, we find the true correlation length also exhibits mean-field behavior with critical exponent $\nu = 0.5$.
That these two \emph{different} correlation lengths obtained from the neural functional are consistent in the critical region, albeit both exhibiting mean-field scaling, gives us confidence in our numerical implementations.

\subsection{Inhomogeneous fluids: predictions for drying, capillary evaporation and local fluctuations}
\label{sec:drying}

\begin{figure}[tbp]
    \centering
    \includegraphics{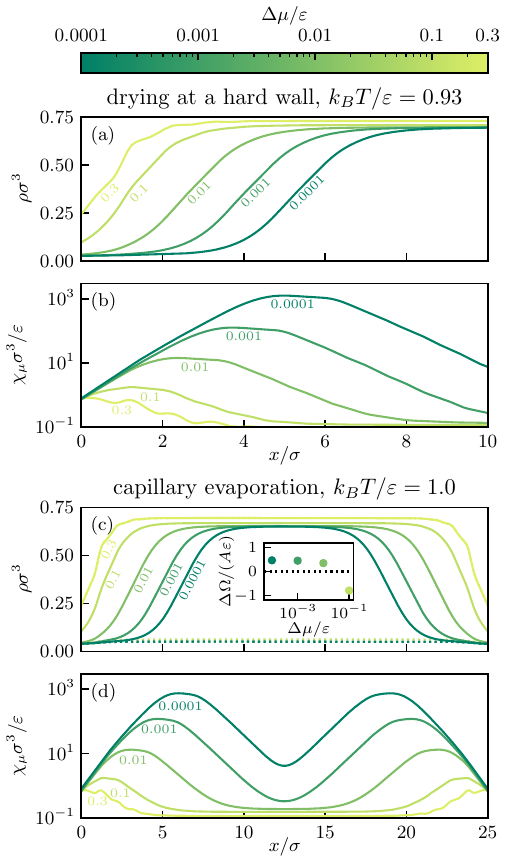}
    \caption{
        Density $\rho(x)$ and local compressibility $\chi_\mu(x)$ profiles for drying at a hard wall (a, b) and for capillary evaporation between hard walls with weak attraction (c, d), cf.\ Eq.~\eqref{eq:Vext_confinement}.
        Results are shown for chemical potentials $\mu = \mu_\mathrm{coex} + \Delta \mu$ approaching, from the liquid side, the respective bulk coexistence values $\mu_\mathrm{coex}$---see text.
        We investigate $\Delta \mu / \varepsilon = 0.3, 0.1, 0.01, 0.001$, and $0.0001$, as labeled.
        In the drying case $k_B T / \varepsilon = 0.93$ and density profiles (a) show a gas layer at the wall whose thickness grows for decreasing values of $\Delta \mu$.
        The local compressibility profile (b) develops a sharp maximum (note the logarithmic scale) at a position located in the emerging gas-liquid interface whose height grows rapidly on approaching bulk coexistence.
        In the case of capillary evaporation $k_B T / \varepsilon = 1.0$ and the density profiles in (c) refer to the condensed `liquid' states (solid lines) and the stable `gas' states (dotted lines) plotted for the same chemical potential differences as in (a).
        The inset displays the difference $\Delta \Omega = \Omega_l - \Omega_g$ between the liquid and gas grand potentials $\Omega_l$ and $\Omega_g$ per lateral system area $A$.
        Positive values of $\Delta \Omega$ indicate metastability of the condensed state.
        Similar to (b), the local compressibility profiles (d) display maxima located in the interface which increase for decreasing $\Delta \mu$.
    }
    \label{fig:drying}
\end{figure}

Previous subsections have focused mainly on bulk behavior, i.e.\ the fluid in the absence of any external potential.
Here we turn to inhomogeneous systems and, in particular, to the physics of adsorption at substrates and of fluids in confinement.
Specifically, we show that the neural functional allows one to capture accurately fluid structure for phenomena associated with phase transitions at state points close to bulk coexistence.
These are challenging to access via simulation studies.
We investigate the phenomena of i) density depletion and drying at a hard-wall and ii) capillary evaporation in a planar slit.
Quantifying the properties associated with such phenomena serves as a stringent test of the neural functional; recall that the training data consist of systems with randomized state points and inhomogeneities not tailored to feature the subtle physics occurring near surface phase transitions.

As before, the equilibrium density profile is calculated via iteration of Eq.~\eqref{eq:EL}, where now a given external potential $V_\mathrm{ext}(\vec{r})$ induces inhomogeneity in $\rho(\vec{r})$.
Examining states close to bulk coexistence requires increased numerical effort, as the equilibrium density profile is highly susceptible to small changes in the control parameters, specifically the deviation of the chemical potential from bulk coexistence.
Nevertheless, the self-consistent calculation via Eq.~\eqref{eq:EL} remains numerically robust, albeit requiring an increasing number of iteration steps.

We consider first the case of a single planar hard wall: $V_\mathrm{ext}(x) = \infty$ for $x < 0$ and $0$ for $x > 0$ and show results in Fig.~\ref{fig:drying}(a).
We choose $k_B T / \varepsilon = 0.93$ and approach bulk coexistence from the liquid side by decreasing the chemical potential $\mu = \mu_\mathrm{coex} + \Delta \mu$ towards the neural prediction $\mu_\mathrm{coex} / \varepsilon \approx -3.55357$ for the chemical potential at bulk coexistence.
The resulting density profiles show the formation of a gas layer adsorbed at the hard wall whose thickness increases continuously when lowering $\Delta \mu$.
This is the classic signature of complete drying, i.e.\ complete wetting of the wall-liquid interface by gas---see e.g.\ Ref.~\cite{Evans1990}, early DFT studies by \citet{Sullivan1981} and \citet{Tarazona1984}, and the important 43 page article by \citet{Henderson1985} that laid out the rich physics involved and presented pioneering simulation results for a square-well model fluid.
This phase transition corresponds to gas layer thickness, or the negative of the Gibbs adsorption, diverging slowly, i.e.\ logarithmically, as $\Delta \mu$ approaches zero.
We note that the shape of the density profile for the smallest value of $\Delta \mu$ plotted resembles closely that of the free interfacial liquid-gas profile shown in Fig.~\ref{fig:tension}(a) for a similar temperature, confirming once again that complete drying occurs.

We next consider confinement between two repulsive hard walls with additional (weak) long-range attraction.
The left wall is described by the external potential
\begin{equation}
    \label{eq:Vext_confinement}
    V_\mathrm{ext}(x) = \begin{cases}
        \infty, &x \leq 0,\\
        \varepsilon_w \left[ \frac{2}{15} \left(\frac{\sigma}{x + x_m}\right)^9 - \left(\frac{\sigma}{x + x_m}\right)^3 \right], &x > 0.
    \end{cases}
\end{equation}
The parameter $\varepsilon_w$ sets the attraction strength and the minimum of the potential well is shifted to $x = 0$ by setting $x_m = (2 / 5)^{1/6} \sigma$.
The right wall is obtained analogously by a mirrored version of Eq.~\eqref{eq:Vext_confinement} and the symmetric confinement potential is the sum of these.
We choose this potential to correspond to a recent MC simulation study by \citet{Wilding2024} and set the attraction strength of the wall $\varepsilon_w = 0.01 \varepsilon$, wall separation $L = 25 \sigma$ and temperature $k_B T / \varepsilon = 1.0$ to be the same as in Ref.~\cite{Wilding2024}.
For such a weak attraction we expect complete drying to occur at a single wall, i.e.\ the corresponding contact angle should be $180^\circ$.
Results for the density profiles are shown in Fig.~\ref{fig:drying}(c) for the same values of $\Delta \mu$ as in panel (a).
For this temperature the neural prediction is $\mu_\mathrm{coex} / \varepsilon \approx -3.46154$.
The density profiles are typical of those found for a fluid in a slit with hard or very weakly attractive walls: a liquid-like density plateau forms at the center of the slit and as $\Delta \mu$ decreases increasing density depletion occurs and eventually a layer of gas develops at the walls.
An early (analytical) DFT study \cite{Tarazona1987}, see also Ref.~\cite{Roth2011}, showed that the `liquid' becomes metastable with respect to the evaporated `gas', whose density is small throughout the slit, and determined the capillary evaporation (first-order) transition point for a model fluid.

Importantly we find that metastability can be investigated using the neural functional.
Recall that Eq.~\eqref{eq:EL} emerges from the minimization \eqref{eq:omegaMinimization}.
For fixed $\Delta \mu$, initialization from an empty system yields the `gas' state, whereas starting with a sufficiently high initial bulk density yields the condensed `liquid' state.
We calculate the grand potential difference $\Delta \Omega = \Omega_l - \Omega_g$ between the `liquid' and `gas' states as a function of chemical potential by evaluating the density functional \eqref{eq:omegaFunctional}, where the excess free energy $F_\mathrm{exc}([\rho], T)$ follows from the functional line integral \eqref{eq:Fexc_funcintegral} employing the previously determined density profiles.
Note that we obtain the grand potential per lateral system area $A$ due to the planar geometry.
We find positive values of $\Delta \Omega / A$ for a chemical potential difference $\Delta \mu / \varepsilon \leq 0.01$---see the inset to Fig.~\ref{fig:drying}(c).
Such states are therefore metastable with respect to capillary evaporation: a gas-like density profile, see the dotted lines in Fig.~\ref{fig:drying}(c), is associated with a lower value of the grand potential.

The one-body density profile provides the crudest measure of the effects of correlations.
We consider also the local compressibility $\chi_\mu(\vec{r}) = \partial \rho(\vec{r}) / \partial \mu$ for fixed $T$ which provides a spatially resolved measure of density fluctuations in the particle number and which was shown to be important in characterizing surface phase transitions \cite{Evans2015,Evans2015a,Evans2017,Eckert2020,Coe2022,Eckert2023,Wilding2024}.
We investigate this quantity for the two cases described above.
Rather than evaluating $\chi_\mu(\vec{r})$ as the partial derivative of the density profile with respect to $\mu$ numerically in terms of a finite difference, we choose to utilize a route employing the fluctuation OZ equation \cite{Eckert2020,Eckert2023}
\begin{equation}
    \label{eq:fluctuation_OZ}
    \chi_\mu(\vec{r}) = \rho(\vec{r}) \int\!\diff{\vec{r}'} c_2(\vec{r}, \vec{r}'; [\rho], T) \chi_\mu(\vec{r}') + \beta \rho(\vec{r}),
\end{equation}
as the inhomogeneous pair direct correlation function $c_2(\vec{r}, \vec{r}'; [\rho], T)$ is explicitly accessible from the neural functional by autodiff, cf.\ Eq.~\eqref{eq:c2}.
Obtaining the local compressibility numerically from Eq.~\eqref{eq:fluctuation_OZ} reduces to solving a system of linear equations and is hence much simpler than the solution of the standard inhomogeneous OZ equation, see e.g.\ Ref.~\cite{Schmidt2022}, owing to the fact that $\chi_\mu(\vec{r})$ is a one-body property.
Such a simplification was already recognized in an early study of wetting transitions \cite{Tarazona1982}.

Panels (b) and (d) in Fig.~\ref{fig:drying} show results for the local compressibility profiles corresponding to the density profiles depicted in panels (a) and (c), respectively.
Approaching bulk coexistence, by decreasing $\Delta \mu$, the local compressibility develops sharp peaks located in the `gas-liquid' interface that emerges as the `gas' layer grows at a wall.
The values of the maxima of $\chi_\mu(x)$ increase by orders of magnitude: note the logarithmic scale.
For the case of drying at a hard wall, see panels (a) and (b), investigated at $k_B T / \varepsilon = 0.93$, well below the temperature $k_B T / \varepsilon \approx 1.04$ where the Fisher-Widom line intersects the binodal, we observe, for the largest $\Delta \mu$, damped oscillations in $\chi_\mu(x)$ reflecting those in the density profile.
For small $\Delta \mu$ we expect from the theory of complete drying, for the potentials we consider here, that the maximum of $\chi_\mu(x)$ should increase as $\Delta \mu^{-1}$ which corresponds to the logarithm of the maximum increasing linearly with the thickness of the drying (gas) layer.
Our results are consistent with this prediction.

The contact values $\chi_\mu(0^+)$ at the hard wall remain almost unchanged as $\Delta \mu$ is reduced.
This behavior is consistent with the contact theorem $\chi_\mu(0^+) = \beta \rho_b$ that establishes an important connection between the bulk density $\rho_b$ far from the wall and a quantity local to the wall \cite{Evans2015,Eckert2020,Eckert2023}.
Our numerical results satisfy this sum rule to within $1\%$, apart from the lowest value of $\Delta \mu$, where numerical errors become more important.
In our investigation of capillary evaporation we chose our system to correspond to that of the MC simulations in Ref.~\cite{Wilding2024} which examined the local compressibility at bulk coexistence.
The density and local compressibility profiles that we obtain mimic those observed in simulation.
However, there are subtleties in making direct comparisons that we return to in Sec.~\ref{sec:discussion_physics}.
We chose to study the profiles at decreasing values of $\Delta \mu$ monitoring the growth of the maximum in the local compressibility which increases in a similar fashion to what is observed for the complete drying case at a single hard wall in Fig.~\ref{fig:drying}(b).

\subsection{Purely supercritical training}
\label{sec:supercritical}

As described in Sec.~\ref{sec:data_generation}, we deliberately trained a second neural network with supercritical data by including only simulations with $k_B T / \varepsilon > 1.3$, see the dotted pink line in Fig.~\ref{fig:overview} showing the temperature cutoff.
In the following we focus on the extrapolation capabilities of this neural network and, in particular, expound how much information can be deduced from learning data at temperatures well above our (subsequent) determination of the critical temperature.

In Fig.~\ref{fig:map_T_rho}, we observe that analyses of the isothermal compressibility $\chi_T$ (first column), true correlation length $\xi$ (second column) and long-range decay of the total correlation function (third column) carry over straightforwardly when using the supercritical neural functional; there are no problems for inferring results at lower temperatures.
Specifically, the predicted \emph{phenomenology} remains identical to the case of training with data including some subcritical states: a spinodal emerges, as characterized by diverging isothermal compressibility and correlation length, $\chi_T^{-1} = 0$ and $\xi^{-1} = 0$.
The top of the spinodal marks a critical point, from which the lines of maximal compressibility and maximal correlation length can be traced and the Fisher-Widom line emerges from the pole analysis of $h(r)$.
Despite requiring extrapolation to substantially lower temperatures than encountered during training, only minor differences occur from the case of full (sub- and supercritical) training.
We note that the critical point is predicted at a slightly higher temperature and density.

Results from purely supercritical training are shown in Fig.~\ref{fig:binodal_comparison} also for the binodal.
Remarkably, the supercritical neural functional in the Euler-Langrange minimization \eqref{eq:EL} delivers at lower temperatures, predicting bulk liquid-gas coexistence.
Moreover, the resulting binodal agrees very well with our previous findings based on the full training data set; small deviations occur for liquid densities at low temperatures.
We emphasize that these predictions of liquid-gas coexistence arise from probing only supercritical states that have no direct signature of possible liquid-gas phase separation.

\section{Discussion}
\label{sec:discussion}

\subsection{Methodology}
\label{sec:discussion_methodology}

Employing the neural functional approach is very different from what one encounters in pure simulation studies or in analytical density functional studies where an explicit approximation to the excess free energy functional is provided from the outset.
Setting up a neural functional from scratch requires an upfront investment in compute resources in order to deliver a suitable training data set.
This (perceived) hurdle is not encountered in working with simulations only; the first run can deliver useful information.
However, once the initial threshold has been passed the method scales very well.
As retraining the neural network itself is relatively cheap, it is entirely feasible and sensible to provide additional training data, be it for changing system size $1/L$, extending parameter ranges, such as temperature $T$, or simply to provide better statistics in an iterative cycle, possibly guided by active machine-learning techniques.
Crucially, the initial investment is never lost, as the original training data can continue to be used in updated training cycles.

What sets aside our neural functional approach is the range of phenomena and results that it can describe: these far exceed the information provided during training.
How this comes about is nontrivial and relies upon the mathematical structure of DFT.
Investigating physical phenomena, i.e.\ calculating the structure and thermodynamic properties,  reduces to standard analysis tasks within the (neural-network-based) functional mapping $c_1(\vec{r}; [\rho], T)$.
Inputting only MC training data of one-body profiles in planar geometry and then examining $c_1(\vec{r}; [\rho], T)$ through the functional lens provides access to quantities which could not be obtained directly from the input data.
Indeed determining these usually requires advanced simulation techniques \cite{Frenkel2023}; note the array of computational methods tailored to study liquid-gas coexistence, interfacial profiles, free energies and the surface tension as well as two-body correlation functions.
These properties and the physical scenarios they encompass follow, in our treatment, from a single numerical object, namely the neural network representing $c_1(\vec{r}; [\rho], T)$.
Our results achieve accuracy comparable with direct simulations of phase coexistence but with much reduced computational cost, and without requiring any additional simulations after training.

We consider our present methodology to be more closely interwoven with the basic theoretical physics at play, in this case DFT, than are the more generic data-based machine-learning methods \cite{Carrasquilla2017,Bedolla2020,Chertenkov2023,Arnold2024}.
Importantly, the formal structure of density functional relationships between generating functionals and hierarchical levels of correlation functions is built into our neural density functional approach and guides application using functional calculus (see Sec.~\ref{sec:neural_theory}).
In this regard, it might be interesting to explore similarities to and differences from other recent work \cite{Carvalho2022,Wu2023,Chen2024} which employs liquid integral equation theory in the context of physics-informed machine learning.

\subsection{Physical phenomena investigated}
\label{sec:discussion_physics}

In this subsection we summarize some of the physics that we considered together with the results we found and raise issues that require further investigation.
For the bulk LJ (continuum) fluid the trained neural direct correlation functional yields an accurate description of pair structure.
The pair correlation function $g(r)$ and static structure factor $S(k)$ exhibit signatures typical for fluids with attractive interactions, see Fig.~\ref{fig:twobody_bulk} (Sec.~\ref{sec:twobody_bulk}).
Moreover, a comprehensive account of \emph{very} subtle correlation phenomena can be obtained with relative computational ease from our determination of the bulk pair direct correlation function.
Specifically we determined the lines in the phase diagram where the isothermal compressibility and true correlation length have their maximal value, and the Fisher-Widom line that denotes the crossover from damped oscillatory to purely monotonic decay of $g(r)$ at large distances, see Fig.~\ref{fig:map_T_rho} (Sec.~\ref{sec:map_T_rho}).
Recall, once again, that the neural network had not encountered bulk pair correlation functions, let alone their asymptotic decay, in training.

Clearly our analysis rests on the familiar (bulk) Ornstein-Zernike (OZ) equation together with the neural representation of the pair direct correlation function $c_2^b(r)$ obtained from automatic differentiation of the (planar) one-body direct correlation functional.
It is tempting to argue that the neural functional provides a highly accurate closure relation for the OZ equation, making stand-alone prediction feasible and avoiding reliance upon approximate closures \cite{Carvalho2022,Wu2023,Chen2024}.
Furthermore, as our approach is DFT-based, there is a unique route to the bulk free energy thereby avoiding inconsistencies that can plague liquid integral equation theories \cite{Hansen2013}.
Note that i) our approach differs from neural functional methods based on pair-correlation matching \cite{Dijkman2024, Sammueller2024pair} and ii) it allows us to access three-body direct correlations, as demonstrated for hard spheres in Ref.~\cite{Sammueller2023neural}.

Our results for the spinodal and the location of the liquid-gas critical point as determined from the divergence of the true correlation length, or equivalently of the compressiblity, are also obtained from $c_2^b(r)$; these are shown in Fig.~\ref{fig:map_T_rho}.
The OZ plots in Fig.~\ref{fig:S_OZ} provide further information about the approach to criticality from above $T_c$: we find mean-field critical exponents for the OZ correlation length and for the compressibility.
Separately we find that the true correlation length diverges in the same fashion as the OZ correlation length.
The spinodal is a concept commonly associated with employing analytic functionals that incorporate attraction, e.g.\ via the mean-field approximation \eqref{eq:FexcMeanField}; of course, in reality there is only a horizontal tie-line in the density-temperature plane linking coexisting gas and liquid.
It seems that our simulation-based machine-learning procedure generates a neural functional, which, for appropriate bulk densities and temperatures, also gives rise to a spinodal.
We examine this further in the next subsection.

In many respects our determination of the gas-liquid binodal (bulk coexistence curve) is one of the most striking results; see Sec.~\ref{sec:coexistence}.
Our training data had no knowledge of bulk phase separation.
Comparison with existing simulation results in Fig.~\ref{fig:binodal_comparison} shows how well the neural prediction performs and how it out-performs the standard analytical DFT.
Using fitting procedures applied previously to simulation data \cite{Panagiotopoulos1994,Wilding1995} for coexisting densities we found evidence for a nonclassical critical exponent $\beta$ and obtained an estimate of $T_c$ that agrees very closely with the best simulation estimate.
The subtleties involved with finite-size effects were discussed in Sec.~\ref{sec:critical_region}.
We have demonstrated that the neural prediction for the binodal is i) highly accurate and ii) internally consistent: results from neural density functional minimization of coexistence states that feature both gas and liquid in a single computational system agree with those from functional line integration for the free energy.
Practical implementation of the latter method requires only a cheap numerical routine; the neural functional is the integrand in Eq.~\eqref{eq:Fexc_funcintegral}.
The bulk equation of state from functional line integration, Eq.~\eqref{eq:eos}, displays a van der Waals loop, see Fig.~\ref{fig:eos}, at subcritical temperatures.
We comment further on this result in Sec.~\ref{sec:discussion_extrapolation}.

The former method of direct numerical stabilization of phase coexistence provides access to one-body liquid-gas interfacial structure.
As remarked in Sec.~\ref{sec:eos_interface_tension}, the density profiles $\rho(x)$ for the two lowest temperatures show the presence of damped oscillations extending into the liquid, see Fig.~\ref{fig:tension}(a).
Such behavior is commonly found in DFT calculations that utilize the standard mean-field approximation \eqref{eq:FexcMeanField}, e.g.\ Ref.~\cite{Evans1993} for the square-well and \citet{Tschopp2020} for the hard-core Yukawa model.
Equivalent behavior is also found in FMT-based DFT calculations for the Asakura-Oosawa model that describes colloid-polymer mixtures.
Pronounced oscillations are found on the (colloid-rich) liquid side of the fluid-fluid interface for states in the neighbourhood of the triple point \cite{Brader2001}.
The physical origin of the oscillations lies, of course, in the packing of particles in the dense liquid.
The argument \cite{Evans1993}, based on asymptotics, is that in the oscillatory region of the bulk phase diagram, as determined by the Fisher-Widom line, the one-body liquid-gas density profile should decay into the bulk liquid at coexistence with the same exponential decay length $1 / \alpha_0$ and same wavelength $2 \pi / \alpha_1$ as would be determined for the bulk state from the pole analysis, i.e.\ Eqs.~\eqref{eq:poles1}--\eqref{eq:poles2_planar}.
Unfortunately there is no simple means of determining the amplitude of oscillations.
Moreover, the argument is based on mean-field ideas and omits effects of thermally induced capillary wave fluctuations.
The physical picture adopted in Ref.~\cite{Evans1993}, and in subsequent work, is that the DFT results provide a `bare' profile that is then `dressed' by unfreezing the fluctuations which serve to erode the oscillations.
For a Gaussian treatment of the fluctuations the decay length and the wavelength of the oscillations are unchanged but the amplitude is reduced by a factor that depends on the interfacial roughness which depends, in turn, upon the interfacial area $A$; see \citet{Brader2003} for a thorough discussion.
The profiles we find here for the LJ fluid exhibit very weak oscillations even at the lowest temperature shown, suggesting that our neural functional might already capture \emph{some} fluctuation effects.
In this context, it is instructive to take note of \citet{Tschopp2020} who show that profiles obtained from a DFT that incorporates correlations into a reference hard-sphere system yields liquid-gas density profiles that have much less pronounced oscillations than those that emerge from employing Eq.~\eqref{eq:FexcMeanField}.
It is tempting to surmise that our neural functional provides a very accurate, albeit still mean-field, description of the `bare' interface.

The liquid-gas surface tension is not beset by subtle issues of including capillary wave fluctuations and we expect our results, obtained from an inhomogeneous functional line integral, to be reliable.
Indeed these match closely high quality simulation data for the LJ system, see Fig.~\ref{fig:tension}(b).
Moreover, our neural functional was able to probe the surface tension for the same range of temperatures as investigated in the calculation of the binodal and we found evidence for a nonclassical critical exponent $\tilde\mu$.

The free liquid-gas interface was the first of our neural density functional investigations of interfacial phenomena.
In Sec.~\ref{sec:drying} we presented results for two further inhomogeneous situations: i) complete drying at a single planar hard wall and ii) capillary evaporation inside a planar slit pore with very weakly attractive walls.
Both constitute demanding tests.
For the former the density and local compressibility profiles must reflect the proper growth of the thickness of the `gas' layer as the chemical potential deviation $\Delta \mu$ is reduced to zero.
Complete drying is a continuous (critical) surface phase transition.
Our neural density functional minimization accounts well for this phenomenon, see Fig.~\ref{fig:drying}(a,b).
In particular we find that the logarithm of the maximum of the local compressibility $\chi_\mu(x)$ increases linearly with the layer thickness,  in agreement with theoretical predictions.
Moreover, we find that the contact theorem for the local compressibility at the hard wall, $\chi_\mu(0^+) = \beta \rho_b$, is satisfied accurately; this provides a valuable check on the consistency of our approach.
In case ii), capillary evaporation, we deliberately chose the confining external potential and temperature to be those employed in the MC study of \citet{Wilding2024}.
Our methods allow us to investigate efficiently several values of the chemical potential deviation $\Delta \mu$ and to measure the grand potential of metastable states thereby allowing us to estimate the value of $\Delta \mu$ at which the (first-order) evaporation transition occurs, see Fig.~\ref{fig:drying}(c,d).
This task is demanding within direct simulation.
For the three smallest values of $\Delta \mu$, corresponding to the three thickest gas layers, the condensed `liquid' state is metastable with respect to the evaporated `gas' state.
For these metastable states $\chi_\mu(x)$ takes on very large values similar to what we find for drying at the planar hard wall in Fig.~\ref{fig:drying}(b).
Note that this quantity does not reach the corresponding bulk value in the middle of the slit; the density fluctuations remain very large throughout.
We find that our results for $\Delta \mu / \varepsilon = 0.01$ match well with the density and local compressibility profiles plotted in Ref.~\cite{Wilding2024}; the maximum of $\chi_\mu(x)$ is about 100 times the bulk value.
However, the results in Ref.~\cite{Wilding2024} refer to $\Delta \mu = 0$, i.e.\ to bulk coexistence as measured in simulation.
This begs, once again, the question as to how best to make comparisons between DFT results and simulation.
Neither the simulation nor the neural functional value of $\mu_\mathrm{coex}$ is known precisely; the former depends on the finite size of the simulation box and the latter on the numerics we employ.
One might argue that for this particular problem the chemical potential itself is a better control parameter than $\Delta \mu$; unlike the case of complete drying there is no critical divergence associated with the approach to coexistence.
We intend to return to this issue in future work.
That we are discussing such subtle matters in this first application of our neural functional to phase transitions attests to the overall potential of our approach.

\subsection{Discovering phase coexistence and extrapolation of functional mappings}
\label{sec:discussion_extrapolation}

The successful reproduction of results with supercritical training only, cf.\ Sec.~\ref{sec:supercritical}, calls for a reassessment of the importance of the one-body direct correlation functional $c_1(\vec{r}; [\rho], T)$; recall this is the quantity we seek to capture from simulation data via a neural network.
It seems feasible to infer this particular functional mapping even when excluding substantial ranges of the parameter space in the training data.
Our findings point to the fundamental nature of the one-body direct correlation functional and its favorable mathematical properties.
It is arguably the object most appropriate for machine-learning tasks---recall also the simplicity of Eq.~\eqref{eq:c1} for determining $c_1(x)$ from simulation data---in contrast to other possible functional mappings \cite{Yatsyshin2022,MalpicaMorales2023}.
Importantly our results show that liquid-gas phase coexistence is an emerging phenomenon that can already be gleaned far from its onset, i.e.\ from training data taken above the critical temperature.

We note in this context that the extrapolation capabilities of the neural functional, and of the underlying functional mapping it represents, are arguably much more extensive than one might initially expect.
Recall that the data used during training, see Sec.~\ref{sec:training}, comprise only true equilibrium states obtained via many-body simulations, and that the resulting density profiles fulfill the Euler-Lagrange equation~\eqref{eq:EL} by construction.
However, when using the trained neural functional for predictions, one cannot guarantee a priori that only such `physical' density profiles are encountered.
This may be the case during the iteration of Eq.~\eqref{eq:EL} as well as for the evaluation of the excess free energy via the parametrized functional integral \eqref{eq:Fexc_funcintegral}.
For the latter the scaled density profiles $\rho_a(\vec{r}) = a \rho(\vec{r})$ are not merely transient, rather they enter as genuine contributions to the total value of $F_\mathrm{exc}([\rho], T)$.

That the predictions remain accurate despite requiring the evaluation of the neural functional with possibly `unphysical' density profiles, i.e.\ ones that cannot correspond to minimization in the presence of any external potential, indicates there is a well-defined continuation of the functional relationship to such states.
Furthermore, it seems feasible to infer this extended mapping from physical data alone, using only the restricted function space of the true equilibrium density profiles that provided input.
We attribute the successful extrapolation to i) the beneficial mathematical properties of the functional mapping $c_1(\vec{r}; [\rho], T)$ and ii) the prowess of the utilized machine-learning techniques.

This is important in practical applications.
Although the evaluation of the neural functional remains valid for `unphysical' density input, the resulting predictions will not correspond directly to what one would find in simulations or in experiments.
For instance, the van der Waals loop in Fig.~\ref{fig:eos} at $T < T_c$ arises from the neural prediction for the pressure corresponding to spatially constant bulk densities $\rho(\vec{r}) = \rho_b$, which are imposed to lie in the metastable or unstable coexistence region.
Clearly, this is not the sequence of density profiles that one encounters in practice where, for fixed mean density, the system will form liquid and gas domains separated by an interface.
If one evaluates the equation of state along this path of physically realizable density profiles, one will indeed find a flat isotherm that is numerically consistent with performing a Maxwell construction in Fig.~\ref{fig:eos}.
Of course, this feature is not unique to the neural functional.
For an analytical DFT, e.g.\ employing the mean-field functional \eqref{eq:FexcMeanField}, one could also solve Eq.~\eqref{eq:EL} for a range of fixed mean densities and determine the resulting density profiles, which will show liquid-gas phase separation within the coexistence region.
These will not be as accurate as the present neural results; leave this aside.
One could then examine the prediction of the theory for the bulk free energy density and calculate the pressure, analogous to Eq.~\eqref{eq:eos}.
Making a Maxwell construction will confirm that the coexisting densities agree with those from solving Eq.~\eqref{eq:EL}.

The ability to analyze the neural functional for (virtually) arbitrary density input is useful and necessary to make certain calculations.
Recall that the determination of the surface tension $\gamma_{lg}$, cf.\ Fig.~\ref{fig:tension}, requires evaluating the functional line integral \eqref{eq:Fexc_funcintegral} along a path of density profiles which crosses the coexistence region.
That our neural results for $\gamma_{lg}$ agree well with simulation serves as an indirect verification of the validity of the neural functional predictions for such unseen inputs.
Similarly, important bulk quantities such as the isothermal compressibility $\chi_T$ and the correlation length $\xi$ can be monitored in states of the bulk fluid that might not be physically accessible, but which provide additional insight---see e.g.\ the emerging spinodal in Fig.~\ref{fig:map_T_rho}.
Of course, this is not conceptually different from using analytical DFT: an explicit approximation for the excess free energy functional will provide these quantities for all states.
What does differ is the mechanism by which the underlying functional is obtained.
Whereas an analytic functional, e.g.\ Eq.~\eqref{eq:FexcMeanField}, is arguably conceived with a certain phenomenology in mind, features such as a spinodal and a van der Waals loop arise for the neural functional solely from training with appropriately chosen simulation data for a given model fluid.

\subsection{Outlook}
\label{sec:discussion_outlook}

There are several problems which should be addressed in future work and which point to possible extensions of the framework we present.

As described in Sec.~\ref{sec:critical_region}, mean-field behavior is found from OZ plots in the close vicinity of the critical point, inline with the construction of the neural functional with a finite box size.
Nevertheless, we envisage that accessing `true' critical behavior is not excluded per se from the neural functional.
The strategy used for the extrapolation to large box sizes via the input node $1/L$ might well prove to be helpful for this purpose.
Supplying additional training data that feature the slow decay of density profiles associated with long-range correlations might also be necessary.
Further, the architecture of the neural network might need to be modified to accommodate the growing correlation length which could require incorporating a larger density window as input to the neural functional.

Although we have focused in this work on the prototypical Lennard-Jones fluid, for ease of comparison with previous accurate simulation data, we see no fundamental problems in applying our methodology to more complex fluids.
As the formal structure of DFT remains intact for arbitrary interaction potentials, the method could be transferred directly to the investigation of systems that feature more elaborate force fields, including three- and higher-body contributions, as used e.g.\ to model interactions in water \cite{Molinero2008,Cisneros2016} or in colloidal gels \cite{Saw2009}.
Methodological extensions arise already for the particular case of pairwise interactions only.
These include the incorporation of bulk pair correlation data during training \cite{Dijkman2024,Sammueller2024pair} and the generalization of the functional dependence to feature explicitly the pair potential $\phi(r)$ \cite{Kampa2024}.
In the future, it would be interesting to investigate further test particle concepts and the topical problem of inverting structural data to infer interparticle interactions, and their uniqueness \cite{Henderson1974}, via neural functional methods.

For the case of anisotropic and molecular fluids, orientational degrees of freedom must be accounted for: the one-body density profile $\rho(\vec{r}, \boldsymbol{\omega})$ which enters the functional mapping must be resolved with respect to both position $\vec{r}$ and orientation $\boldsymbol{\omega}$.
In a first venture, \citet{Simon2024patchy} demonstrated that machine learning a classical density functional is feasible for the anisotropic Kern-Frenkel model in planar geometry.
Notably, the authors found that the external flat-wall potentials for the generation of training data had to incorporate nontrivial orientation-dependence in order to probe the functional mapping sufficiently.
Another very recent work considers an application to a model of carbon dioxide, where orientations become important \cite{Yang2024}.
Methods developed in the context of molecular DFT \cite{Jeanmairet2013,Ding2017} could serve as a practical guide on how to deal with numerical challenges arising from the additional orientational resolution; see the recent Ref.~\cite{Simon2024anisotropic}.

For mixtures, the relevant functionals depend on the density profiles $\rho_i(\vec{r})$ of the individual species $i = 1, \dots, s$.
A machine-learning scheme aiming to represent the direct correlation functional $c_{1, i}(\vec{r}; [\rho_1, \dots, \rho_s], T)$ must therefore account for the additional species labeling.
Tackling the fluid phase behavior of multi-component systems is, of course, important in physical chemistry and chemical engineering.
Note, for example, that rich phase behavior emerges already for the case of a very simple binary mixture \cite{Wilding1998}.
In ionic systems, charge ordering is key and appropriate number-number and charge-charge correlation functions should be distinguished.
For the restricted primitive model, \citet{Bui2024} have successfully demonstrated training a neural density functional, taking into account ideas from local molecular field theory to curb the long-range electrostatic interactions.

We have shown that employing only planar geometry in the input is sufficient to predict various aspects of liquid-gas phase coexistence.
Crystallization presents an even bigger challenge.
Describing the freezing transition is likely to require input with knowledge of the three-dimensional geometry.
Equivariant neural networks \cite{Cohen2016,Satorras2021}, which implement certain symmetry conditions directly in the neural network architecture, could be beneficial in this case to ensure efficient data utilization and performant inference.

As a final remark we point out that describing interacting many-body systems in terms of functional relationships is not only relevant for classical statistical physics and systems in thermodynamic equilibrium.
Electronic DFT, specifically in its Kohn-Sham formulation, plays a role completely analagous to that of classical DFT.
It provides an exact formulation of the many-body quantum mechanical treatment of interacting electrons and, of course, constitutes a cornerstone in modern computational chemistry and condensed matter physics \cite{Burke2012}.
The formal similarities are deep: the electron density $n(\vec{r})$ takes on the role of $\rho(\vec{r})$ as functional input, with the central functionals now being the exchange-correlation energy functional $E_\mathrm{xc}[n]$ and the exchange-correlation potential $v_\mathrm{xc}(\vec{r}; [n])$, which is generated from $E_\mathrm{xc}[n]$ via functional differentiation.
Our findings in the classical case, where the direct correlation functional $c_1(\vec{r}; [\rho], T)$ is generated from the excess free energy functional $F_\mathrm{exc}([\rho], T)$, cf.\ Eq.~\eqref{eq:c1_from_Fexc}, immediately suggest representing $v_\mathrm{xc}(\vec{r}; [n])$ as a neural functional and then taking inspiration from the functional calculus methods described in Sec.~\ref{sec:neural_theory}.

While classical and quantum DFT, as described, operate strictly in thermodynamic equilibrium, power functional theory \cite{Schmidt2013,Schmidt2022} establishes a formally exact extension to dynamics and nonequilibrium systems.
The pertinent functional relationships must be augmented.
For the case of classical overdamped Brownian motion, besides the density profile, the entire history of the one-body current $\vec{J}(\vec{r}, t)$ enters the functional mapping, which is now formulated in terms of the internal force $\vec{f}_\mathrm{int}(\vec{r}, t; [\rho, \vec{J}])$.
For steady states, where dependence on time $t$ vanishes, constructing a neural force functional has been demonstrated successfully \cite{Heras2023,Zimmermann2024}; this offers a promising perspective for further generalizations.

\section*{Data availability}

The data that support the findings of this study are openly available in Zenodo \cite{Zenodo}.

\begin{acknowledgments}
We thank Francesco Turci and Nigel Wilding for helpful discussions.
R.~E.\ acknowledges support of the Leverhulme Trust, grant no.\ EM-2020-029/4.
This work is supported by the DFG (Deutsche Forschungsgemeinschaft) under project nos.\ 551294732 and 422127126.
\end{acknowledgments}

\bibliography{bibliography.bib}

\end{document}